# General vector auxiliary differential equation finite-difference time-domain method for rotationally symmetric vector wave propagation in nonlinear optics

**CALEB J. GRIMMS*** **AND ROBERT D. NEVELS**

*Department of Electrical and Computer Engineering, Texas A&M University, 188 Bizzel St, College Station, TX 77801, USA*
**cgrimms@tamu.org*

**Abstract:** In this paper the theory and simulation results are presented for 3D vector cylindrical rotationally symmetric electromagnetic wave propagation in an isotropic nonlinear medium using a modified finite-difference time-domain general vector auxiliary differential equation method for nonlinear polarization vectors, the nonlinear dipole moment per unit volume, including both an isotropic part and an anisotropic part. The theory is presented for both the transverse magnetic and transverse electric cases along with the combined equations. The simulation results for transverse electric spatial soliton propagation in BK7, and the simulation results for transverse electric and transverse magnetic spatial soliton propagation in fused silica, assuming a purely isotropic polarization vector, are shown. The simulation results for both transverse magnetic and transverse electric spatial soliton propagation in carbon disulfide, including the anisotropic part of the polarization vector, are also shown.

## 1. Introduction

In this paper we present the simulation of vector cylindrical rotationally symmetric (CRS) electromagnetic wave propagation in nonlinear dispersive material at optical frequencies using the general vector auxiliary differential equation finite-difference time-domain (FDTD GVADE) method [1-5] extending and generalizing the 3D CRS FDTD GVADE method [6].

The work on vector cylindrical rotationally symmetric (CRS) optical waves propagating in nonlinear materials began in the early 1970's with Dieter Pohl's work [7-9] and the work of others such as [10], where CRS waves in nonlinear media were first analyzed and later experimentally produced in carbon disulfide ($CS_2$). Pohl's theoretical work in [7] mostly followed a similar approach to that used in earlier work on scalar cylindrical waves [11, 12], modifying and expanding it for vector analysis. Part of Pohl's conclusion in [7] was that the longitudinal component of the electric field played a role in the wave propagation for the transverse magnetic to z ($TM_z$) case and could not be neglected in some circumstances, as is frequently done in optical field analysis. He showed that for $TM_z$ waves, where a longitudinal electric field is present, there were cases where the results diverged from the transverse electric to z ($TE_z$) case, where there is no longitudinal electric field component. However, for many cases, the $TM_z$ and $TE_z$ results were indistinguishable from each other, implying the longitudinal field may be neglected at times. Another part of [7] was to state that the nonlinear polarization vector couples the $TM_z$ electric field wave equation's *r* and *z* vector components together, due to the tensor relationship between the polarization vector and the electric field.

Part of this paper's goal is to numerically simulate 3D CRS waves, reproducing similar 3D CRS waves to Pohl's experiments in $CS_2$ [7]. However, to properly simulate 3D CRS waves similar to Pohl's experiment in $CS_2$, the 3D CRS FDTD GVADE method [6] was extended to include an anisotropic part of the polarizations vector as discussed in [5], since $CS_2$ is an isotropic media with a nonlinear polarization vector with a larger anisotropic part than isotropic part, unlike other material with weaker or negligible anisotropic components. Fused silica for example, has a weaker anisotropic part relative to its isotropic part than $CS_2$ at optical frequencies, which may be ignored in some cases [5], without significant difference in the result, though a difference will be present.

One of the primary ways optical waves have been analyzed, is using the Nonlinear Schrödinger Equation (NLSE) [13, 14], which is a simplification of the vector wave equation [15, 16]. It is classically solved using the Split-Step Fourier Method (SSFM). The rotationally symmetric problem has been solved using this method as well [13, 17, 18]. The approach of using the SSFM to solve the NLSE is a well a developed technique and it works well when the underlying assumptions are met, the paraxial approximation and the slowly varying envelope approximation [16] for example, but these approximations limit the types of problems that can be solved using the technique. For another example, in its most basic form, the SSFM is unable to solve reflection problems, since it assumes the wave only propagates in one direction.

Another way that optical waves have been analyzed, is the FDTD GVADE method [1-5], which is extended by the 3D CRS FDTD GVADE method [6]. The FDTD GVADE method models electromagnetic wave propagation at optical frequencies in a nonlinear material for linear and nonlinear frequency dependent material response, using auxiliary differential equations (ADEs) as described on page 498 of [4], "as a means to extend FDTD to incorporate polarization by time-stepping auxiliary differential equations concurrently with Maxwell's curl equations". The historical development of extending the FDTD method by using ADEs is summarized well in [4], and a few selected historical papers with more details on ADE extensions of the FDTD method and their development are [1-5, 19-24]. It should be noted that ADE extensions to FDTD methods have also been used for electromagnetic wave propagation in linear dispersive anisotropic media [25]; however, this paper is discussing electromagnetic wave propagation in purely isotropic nonlinear media with a polarization vector including both isotropic and anisotropic parts, which is different than modeling electromagnetic wave propagation in an anisotropic material. Like all numerical methods, the 3D CRS FDTD GVADE method has limitations as well. For example, larger computational resources are generally required to solve problems using the FDTD GVADE method as compared with solving the same problems with the NLSE using the SSFM. An advantage of the 3D CRS FDTD GVADE method compared to the 3D Cartesian FDTD GVADE method, is the 3D CRS method simulates a 3D wave using 2D computational resources, by taking advantage of rotational symmetry, mitigating some of the computational resource limitations of the FDTD GVADE method. Since the FDTD GVADE method directly solves Maxwell's Equations, it can avoid some of the approximations used in the NLSE, allowing some problems to be solved more accurately.

## 2. Theory

The theory behind the generalized 3D CRS FDTD GVADE method is presented in the following sections. The theory is an extension and combination of the previous work [6] and [5]. First, Maxwell's equations are reviewed for the vector cylindrical rotationally symmetric case. Second, the nonlinear polarization vector is discussed. Third, the FDTD GVADE method is presented with the polarization vector including an isotropic and an anisotropic part, which is required for the carbon disulfide example. Lastly, the Newton-Raphson method is presented for the 3D CRS FDTD GVADE method.

### *2.1 Rotationally Symmetric Maxwell's Equations*

Maxwell's equations for a nonmagnetic material without any free charges $\rho$ or impressed current density vector $\mathbf{J}$ are:

$$\nabla \times \mathbf{H} = \frac{\partial \mathbf{D}}{\partial t} \,, \tag{1}$$

$$\nabla \times \mathbf{E} = -\mu_0 \frac{\partial \mathbf{H}}{\partial t} \,, \tag{2}$$

where **H** is the instantaneous magnetic field vector and **D** is the electric flux density (otherwise known as the electric displacement vector). The Ampère-Maxwell law can be re-written in terms of the instantaneous electric field vector **E**, the induced polarization current density vector **J** and induced polarization density vector **P** defined as the "electric dipole moment per unit volume", where $\mathbf{D} = \varepsilon_0\mathbf{E} + \mathbf{P}$ and $\mathbf{J} = \frac{\partial \mathbf{P}}{\partial t}$ [4, 15, 26]:

$$\nabla \times \mathbf{H} = \varepsilon_0 \frac{\partial \mathbf{E}}{\partial t} + \mathbf{J} \,. \tag{3}$$

Writing the equations in cylindrical coordinates [27],

$$\nabla \times \mathbf{H}(r, \phi, z) = \left[\frac{1}{r}\frac{\partial H_z}{\partial \phi} - \frac{\partial H_\phi}{\partial z},\ \frac{\partial H_r}{\partial z} - \frac{\partial H_z}{\partial r},\ \frac{1}{r}\frac{\partial}{\partial r}(rH_\phi) - \frac{1}{r}\frac{\partial H_r}{\partial \phi}\right]^{\mathrm{T}} = \varepsilon_0\left[\frac{\partial E_r}{\partial t}, \frac{\partial E_\phi}{\partial t}, \frac{\partial E_z}{\partial t}\right]^{\mathrm{T}} + \mathbf{J} \,, \tag{4}$$

$$\nabla \times \mathbf{E}(r, \phi, z) = \left[\frac{1}{r}\frac{\partial E_z}{\partial \phi} - \frac{\partial E_\phi}{\partial z},\ \frac{\partial E_r}{\partial z} - \frac{\partial E_z}{\partial r},\ \frac{1}{r}\frac{\partial}{\partial r}(rE_\phi) - \frac{1}{r}\frac{\partial E_r}{\partial \phi}\right]^{\mathrm{T}} = -\mu_0\left[\frac{\partial H_r}{\partial t}, \frac{\partial H_\phi}{\partial t}, \frac{\partial H_z}{\partial t}\right]^{\mathrm{T}} . \tag{5}$$

Then, for the 3D rotationally symmetric case, assuming vector rotational symmetry results in all $\frac{\partial}{\partial \phi}$ terms being zero due to the fields being constant in the $\phi$-dimension/direction.

$$\nabla \times \mathbf{H}(r, \phi, z) = \left[-\frac{\partial H_\phi}{\partial z},\ \frac{\partial H_r}{\partial z} - \frac{\partial H_z}{\partial r},\ \frac{1}{r}\frac{\partial}{\partial r}(rH_\phi)\right]^{\mathrm{T}} = \varepsilon_0\left[\frac{\partial E_r}{\partial t}, \frac{\partial E_\phi}{\partial t}, \frac{\partial E_z}{\partial t}\right]^{\mathrm{T}} + \mathbf{J} \,, \tag{6}$$

$$\nabla \times \mathbf{E}(r, \phi, z) = \left[-\frac{\partial E_\phi}{\partial z},\ \frac{\partial E_r}{\partial z} - \frac{\partial E_z}{\partial r},\ \frac{1}{r}\frac{\partial}{\partial r}(rE_\phi)\right]^{\mathrm{T}} = -\mu_0\left[\frac{\partial H_r}{\partial t}, \frac{\partial H_\phi}{\partial t}, \frac{\partial H_z}{\partial t}\right]^{\mathrm{T}} . \tag{7}$$

The cylindrical rotationally symmetric fields can be split into $\mathrm{TM}_z$ and $\mathrm{TE}_z$ parts. Writing out the transverse magnetic to $z$, $\mathrm{TM}_z$, equations by grouping the vector component equations in equations (6) and (7) only having magnetic fields in the transverse $\phi$ direction:

$$(\nabla \times \mathbf{H})_{r,\mathrm{TM}} = -\frac{\partial H_\phi}{\partial z} = \varepsilon_0 \frac{\partial E_r}{\partial t} + J_r \,, \tag{8}$$

$$(\nabla \times \mathbf{H})_{z,\mathrm{TM}} = \frac{1}{r}\frac{\partial}{\partial r}\left(rH_\phi\right) = \varepsilon_0 \frac{\partial E_z}{\partial t} + J_z \,, \tag{9}$$

$$(\nabla \times \mathbf{E})_{\phi,\mathrm{TM}} = \frac{\partial E_r}{\partial z} - \frac{\partial E_z}{\partial r} = -\mu_0 \frac{\partial H_\phi}{\partial t} . \tag{10}$$

Writing out the transverse electric to $z$, $\mathrm{TE}_z$, equations by grouping the vector component equations in equations (6) and (7) only having electric fields in the transverse $\phi$ direction:

$$(\nabla \times \mathbf{H})_{\phi,\mathrm{TE}} = \frac{\partial H_r}{\partial z} - \frac{\partial H_z}{\partial r} = \varepsilon_0 \frac{\partial E_\phi}{\partial t} + J_\phi \,, \tag{11}$$

$$(\nabla \times \mathbf{E})_{r,\mathrm{TE}} = \frac{\partial E_\phi}{\partial z} = \mu_0 \frac{\partial H_r}{\partial t} \,, \tag{12}$$

$$(\nabla \times \mathbf{E})_{z,\mathrm{TE}} = \frac{1}{r}\frac{\partial}{\partial r}\left(rE_{\phi}\right) = -\mu_0 \frac{\partial H_z}{\partial t}. \tag{13}$$

The $TE_z$ and $TM_z$ equations are independent of each other in linear media, but for nonlinear media, when both $TE_z$ and $TM_z$ fields are present, the nonlinear polarization current terms connect the equations, through the nonlinear components of the polarization vector [7, 10].

### *2.2 Polarization Vector and Susceptibility Tensors*

The polarization vector for nonlinear optics in many materials can be written in terms of a linear $\mathbf{P}^{(1)}$ component and nonlinear third order component $\mathbf{P}^{(3)}$, assuming the $\mathbf{P}^{(2)}$ and higher order terms are insignificant,

$$\mathbf{P}(\mathbf{r},t) = \mathbf{P}^{(1)}(\mathbf{r},t) + \mathbf{P}^{(3)}(\mathbf{r},t). \tag{14}$$

For many materials at optical frequencies, the third order component can be written in a simplified form using the Born-Oppenheimer approximation, where the "electronic" and "nuclear" contribution of $\mathbf{P}^{(3)}(\mathbf{r},t)$ are separated as,

$$\mathbf{P}^{\mathrm{NL}}(\mathbf{r},t) = \varepsilon_0 \boldsymbol{\chi}_{\mathrm{el}}^{(3)} \mathbf{E}(\mathbf{r},t)|\mathbf{E}(\mathbf{r},t)|^2 + \varepsilon_0 \mathbf{E}(\mathbf{r},t) \int_{-\infty}^{\infty} \left[ \boldsymbol{\chi}_{\mathrm{nu}}^{(3)}(t-t')\vdots\mathbf{E}(\mathbf{r},t')\mathbf{E}(\mathbf{r},t')\right] dt' \tag{15}$$

where the symbol "$\vdots$" represents a tensor product, $\boldsymbol{\chi}_{\mathrm{el}}^{(3)}$ is the is electronic susceptibility and $\boldsymbol{\chi}_{\mathrm{nu}}^{(3)}$ is the rank 4 nuclear susceptibility tensor. Please see Supplement 1 (Section 1: The Polarization Vector and Rank 4 Susceptibility Tensor in Cartesian and Cylindrical Coordinates Using Index Notation) for more discussion of the polarization vector and the susceptibility tensor in cylindrical coordinates. The nuclear integral can also be written in vector form without the tensor product as [15, 28, 29]:

$$\begin{aligned} \mathbf{P}_{\mathrm{nu}}^{\mathrm{NL}} &= \varepsilon_0 \, \mathbf{E}(\mathbf{r},t) \int_{-\infty}^{\infty} \left[ \chi_{\mathrm{nu,a}}^{(3)}(t-t') \, |\mathbf{E}(\mathbf{r},t')|^2 \right] dt' + \\ &\varepsilon_0 \int_{-\infty}^{\infty} \left[ [\mathbf{E}(\mathbf{r},t) \cdot \mathbf{E}(\mathbf{r},t')] \;\; \chi_{\mathrm{nu,b}}^{(3)}(t-t') \, \mathbf{E}(\mathbf{r},t') \right] dt', \end{aligned} \tag{16}$$

where "a" and "b" correspond to the isotropic and anisotropic parts respectively.

### *2.3 Extended FDTD GVADE Method Summary: Isotropic Media with a Polarization Vector Including Both Isotropic and Anisotropic Parts*

The extended FDTD GVADE method [1, 2, 4, 5] models electromagnetic wave propagation in nonlinear isotropic media at optical frequencies by accounting for the nonlinear behavior through the polarization current terms in Ampère's law. The Ampère's law equation is written at the time-step $n + 1/2$ approximating the time derivative by using a finite-difference centered at $n + 1/2$, and is solved for the electric fields at the time-step $n + 1$ using the multi-dimensional Newton-Raphson method:

$$\nabla \times \mathbf{H}^{n+1/2} = \frac{\varepsilon_0}{\Delta t}(\mathbf{E}^{n+1} - \mathbf{E}^{n}) + \mathbf{J}_{\mathrm{Lorentz}}^{n+1/2} + \mathbf{J}_{\mathrm{el}}^{n+1/2} + \mathbf{J}_{\mathrm{nu,a}}^{n+1/2} + \mathbf{J}_{\mathrm{nu,b}}^{n+1/2}. \tag{17}$$

Then, using the electric fields at time-step $n + 1$ as the new electric fields at time-step $n$, Faraday's law is solved for the magnetic fields using classic finite-difference time-domain update equations at the new time-step $n + 1/2$. This is repeated for each new time-step $n$.

The method solves the convolution integrals from equation (16), where $\mathbf{J} = \frac{\partial \boldsymbol{P}}{\partial t}$, as "auxiliary differential equations", ADEs, which are solved as finite-difference update equations in Ampère's law using the Newton-Raphson method. The isotropic and anisotropic convolution integrals defined as a scalar auxiliary variable are

$$S_{\text{nu,a}}(t) = \int_{-\infty}^{\infty} \left[\chi^{(3)}_{\text{nu,a}}(t-t')\,|\mathbf{E}(\mathbf{r},t')|^2\right] dt' = \chi_0^{(3)} f_{\text{nu,a}} \left(g_{\text{nu,a}}(t) * |\mathbf{E}(t)|^2\right) , \tag{18}$$

$$S_{\text{nu,b},kl}(t) = \int_{-\infty}^{\infty} \left[\chi^{(3)}_{\text{nu,b}}(t-t')\; E_k(t')E_l(t')\right] dt' = \chi_0^{(3)} f_{\text{nu,b}} \left(g_{\text{nu,b}}(t) * [E_k(t)E_l(t)]\right) . \tag{19}$$

After choosing $g_{\text{nu,a}}(t)$ and $g_{\text{nu,b}}(t)$ that approximate/model the nonlinear behavior, where each have "closed form" analytical Fourier Transforms, the auxiliary differential equation is created by taking a Fourier Transform of equations (18) and (19), then algebraically re-arranged. The auxiliary differential equation is solved by taking the Inverse Fourier Transform, and then a finite central difference approximation centered around time-step $n$ is used to approximate the ADEs. Lastly, the ADEs are solved for the scalar auxiliary variables at the new time-step $n + 1$ as update equations. The polarization current is then found by numerically differentiating $\mathbf{J} = \frac{\partial \mathbf{P}}{\partial t}$ using a central difference centered around time-step $n + 1/2$. The Extended FDTD GVADE method [1, 2, 4, 5] solves for the linear and nonlinear polarization current vectors using,

$$\mathbf{J}^{n+1/2}_{\text{Lorentz}} = \frac{1}{2}\sum_{p=1}^{3}\left[\left(1+\alpha_p\right)\mathbf{J}^{n}_{\text{Lorentz}_p} - \mathbf{J}^{n-1}_{\text{Lorentz}_p} + \frac{\gamma_p}{2\Delta t}\left(\mathbf{E}^{n+1} - \mathbf{E}^{n-1}\right)\right], \tag{20}$$

$$\mathbf{J}^{n+1/2}_{\text{el}} = \chi_0^{(3)} f_{\text{el}} \frac{\varepsilon_0}{\Delta t}\left\{\left(\left|\mathbf{E}^{n+1}\right|\right)^2 \mathbf{E}^{n+1} - (|\mathbf{E}^n|)^2\mathbf{E}^n\right\}, \tag{21}$$

$$\mathbf{J}^{n+1/2}_{\text{nu,a}} = \frac{\varepsilon_0}{\Delta t}\left(\mathbf{E}^{n+1} S^{n+1}_{\text{nu,a}} - \mathbf{E}^{n} S^{n}_{\text{nu,a}}\right), \tag{22}$$

$$\mathbf{J}^{n+1/2}_{\text{nu,b}} = \frac{\varepsilon_0}{\Delta t}\left(\mathbf{S}^{n+1}_{\text{nu,b}}\mathbf{E}^{n+1} - \mathbf{S}^{n}_{\text{nu,b}}\mathbf{E}^{n}\right), \tag{23}$$

where,

$$\mathbf{P}^{(3)}_{\text{nu,b}}(t) = \varepsilon_0 \mathbf{S}_{\text{nu,b}}\mathbf{E}(t) = \varepsilon_0 \begin{bmatrix} S_{\text{nu,b},rr}(t) & S_{\text{nu,b},r\phi}(t) & S_{\text{nu,b},rz}(t) \\ S_{\text{nu,b},r\phi}(t) & S_{\text{nu,b},\phi\phi}(t) & S_{\text{nu,b},\phi z}(t) \\ S_{\text{nu,b},rz}(t) & S_{\text{nu,b},\phi z}(t) & S_{\text{nu,b},zz}(t) \end{bmatrix} \begin{bmatrix} E_r(t) \\ E_\phi(t) \\ E_z(t) \end{bmatrix}, \tag{24}$$

$$J^{n+1/2}_{\text{nu,b},k} = \frac{\varepsilon_0}{\Delta t}\left[E_r^{n+1} S^{n+1}_{\text{nu,b},kr} - E_r^{n} S^{n}_{\text{nu,b},kr}\right] + \frac{\varepsilon_0}{\Delta t}\left[E_\phi^{n+1} S^{n+1}_{\text{nu,b},k\phi} - E_\phi^{n} S^{n}_{\text{nu,b},k\phi}\right] + \frac{\varepsilon_0}{\Delta t}\left[E_z^{n+1} S^{n+1}_{\text{nu,b},kz} - E_z^{n} S^{n}_{\text{nu,b},kz}\right]. \tag{25}$$

It should be noted that $S_{\text{nu,b},r\phi}(t) = S_{\text{nu,b},\phi r}(t)$, $S_{\text{nu,b},rz}(t) = S_{\text{nu,b},zr}(t)$ and $S_{\text{nu,b},\phi z}(t) = S_{\text{nu,b},z\phi}(t)$. Please see [4] and [5] for the full derivation and more information.

### *2.4 Combined $TE_z$ and $TM_z$ Equations and Solution*

The general rotationally symmetric electromagnetic wave is the combination of the $TE_z$ and $TM_z$ waves, and in the case of linear media these equations can be solved independently if both are present, but in the case of nonlinear media, the Ampère's law vector component equations are coupled nonlinear equations [10], requiring the electric field components to be determined using the 3D Newton-Raphson method. Following the approach from [1, 2, 4-6], the electric field at time-step $n$ + 1 is determined using the Newton-Raphson method to solve Ampère's law at time-step $n$ + 1/2 from equation (17). Plugging in the polarization current expressions into Ampère's law at time step $n$ + 1/2, results in an equation which can be used as part of the Newton-Raphson method to determine $\mathbf{E}^{n+1}$, where $\mathbf{E}^{n+1} = \hat{\boldsymbol{a}}_r E_r^{n+1} + \hat{\boldsymbol{a}}_\phi E_\phi^{n+1} + \hat{\boldsymbol{a}}_z E_z^{n+1}$,

$$\begin{bmatrix} R \\ \Phi \\ Z \end{bmatrix} = -\nabla\times\mathbf{H}^{n+1/2} + \frac{\varepsilon_0}{\Delta t}\left(\mathbf{E}^{n+1} - \mathbf{E}^{n}\right) + \tag{26}$$
$$\frac{1}{2}\sum_{p=1}^{3}\left[(1+\alpha_p)\mathbf{J}^{n}_{\mathrm{Lorentz}_p} - \mathbf{J}^{n-1}_{\mathrm{Lorentz}_p} + \frac{\gamma_p}{2\Delta t}\left(\mathbf{E}^{n+1} - \mathbf{E}^{n-1}\right)\right] +$$
$$\chi_0^{(3)} f_{\mathrm{el}} \frac{\varepsilon_0}{\Delta t}\left\{\left(|\mathbf{E}^{n+1}|\right)^2\mathbf{E}^{n+1} - (|\mathbf{E}^{n}|)^2\mathbf{E}^{n}\right\} + \frac{\varepsilon_0}{\Delta t}\left(\mathbf{E}^{n+1}S^{n+1}_{\mathrm{nu,a}} - \mathbf{E}^{n}S^{n}_{\mathrm{nu,a}}\right) +$$
$$\frac{\varepsilon_0}{\Delta t}\left(\mathbf{S}^{n+1}_{\mathrm{nu,b}}\mathbf{E}^{n+1} - \mathbf{S}^{n}_{\mathrm{nu,b}}\mathbf{E}^{n}\right).$$

The Newton-Raphson method solves for the electric field by iterative guessing based on the Jacobian Matrix, until $R$, $\Phi$ and $Z$ become sufficiently close to zero, using [30]

$$\begin{bmatrix} E_r^{n+1} \\ E_\phi^{n+1} \\ E_z^{n+1} \end{bmatrix}_{g+1} = \begin{bmatrix} E_r^{n+1} \\ E_\phi^{n+1} \\ E_z^{n+1} \end{bmatrix}_{g} - \left(\mathbf{M}^{-1}\begin{bmatrix} R \\ \Phi \\ Z \end{bmatrix}\right)\Bigg|_g . \tag{27}$$

Each Newton-Raphson iteration guess values of $E_r^{n+1}$, $E_\phi^{n+1}$ and $E_z^{n+1}$ at guess number "$g$", are used to then estimate the next iteration values of $E_r^{n+1}$, $E_\phi^{n+1}$ and $E_z^{n+1}$ at guess number "$g$+1". The Jacobian matrix is defined by $\partial(R,\Phi,Z)/\partial(E_r^{n+1},E_\phi^{n+1},E_z^{n+1})$,

$$M_{11} = \frac{\varepsilon_0}{\Delta t}\left[1 + f_{\mathrm{el}}\chi_0^{(3)}\left\{3\left(E_r^{n+1}\right)^2 + \left(E_\phi^{n+1}\right)^2 + \left(E_z^{n+1}\right)^2\right\} + S^{n+1}_{\mathrm{nu,a}} + S^{n+1}_{\mathrm{nu,b},rr}\right] + \frac{1}{4\Delta t}(\gamma_1 + \gamma_2 + \gamma_3) , \tag{28}$$

$$M_{22} = \frac{\varepsilon_0}{\Delta t}\left[1 + f_{\mathrm{el}}\chi_0^{(3)}\left\{\left(E_r^{n+1}\right)^2 + 3\left(E_\phi^{n+1}\right)^2 + \left(E_z^{n+1}\right)^2\right\} + S^{n+1}_{\mathrm{nu,a}} + S^{n+1}_{\mathrm{nu,b},\phi\phi}\right] + \frac{1}{4\Delta t}(\gamma_1 + \gamma_2 + \gamma_3) , \tag{29}$$

$$M_{33} = \frac{\varepsilon_0}{\Delta t}\left[1 + f_{\mathrm{el}}\chi_0^{(3)}\left\{\left(E_r^{n+1}\right)^2 + \left(E_\phi^{n+1}\right)^2 + 3\left(E_z^{n+1}\right)^2\right\} + S^{n+1}_{\mathrm{nu,a}} + S^{n+1}_{\mathrm{nu,b},zz}\right] + \frac{1}{4\Delta t}(\gamma_1 + \gamma_2 + \gamma_3) , \tag{30}$$

$$M_{12} = M_{21} = f_{\mathrm{el}}\frac{\varepsilon_0\chi_0^{(3)}}{\Delta t}2E_r^{n+1}E_\phi^{n+1} + \frac{\varepsilon_0}{\Delta t}(S^{n+1}_{\mathrm{nu,b},r\phi}) , \tag{31}$$

$$M_{13} = M_{31} = f_{el}\frac{\varepsilon_0\chi_0^{(3)}}{\Delta t}2E_r^{n+1}E_z^{n+1} + \frac{\varepsilon_0}{\Delta t}(S^{n+1}_{\mathrm{nu,b},rz}) , \tag{32}$$

$$M_{23} = M_{32} = f_{el}\frac{\varepsilon_0\chi_0^{(3)}}{\Delta t}2E_\phi^{n+1}E_z^{n+1} + \frac{\varepsilon_0}{\Delta t}(S^{n+1}_{\mathrm{nu,b},\phi z}) . \tag{33}$$

Note that these equations have the same form as the Cartesian equations [5], replacing $x$ with $r$ and $y$ with $\phi$. However, the $(\nabla\times\mathbf{H})_{z,\mathrm{TM}} = (1/r)\partial/\partial r\left(rH_\phi\right)$ used in the Newton-Raphson method and the $(\nabla\times\mathbf{E})_{z,\mathrm{TE}} = (1/r)\partial/\partial r\left(rE_\phi\right)$ equation used as a finite-difference update equation to determine $H_z$, are different than the Cartesian case, where the 2D $\mathrm{TM}_z$ and $\mathrm{TE}_z$ equations have the form of $(\nabla\times\mathbf{H})_{z,\mathrm{TM,2D}} = \partial/\partial x\left(H_y\right)$ and $(\nabla\times\mathbf{E})_{z,\mathrm{TE,2D}} = \partial/\partial x\left(E_y\right)$.

Another difference between the rotationally symmetric solution and the Cartesian solution, is the general solution of the general 3D rotationally symmetric fields can be solved using a 2D numerical grid, requiring 2D computational resources, while simulating the same 3D wave using the general 3D solution, a 3D numerical grid and 3D computational resources are required. For this paper, the combined isotropic polarization vector case was chosen as an example in Section 4.2, meaning the terms corresponding to $\mathbf{S}_{\text{nu,b}}$ were removed from the above equations, due to the extensive computational resources required for simulating the general 3D CRS case where the polarization vector includes both the isotropic and anisotropic parts.

### *2.5 TE$_z$ Equations*

For the special case of transverse to z electric fields, the nonlinear polarization vector $\mathbf{P}^{(3)}(t)$ simplifies greatly, removing the anisotropic behavior,

$$P_\phi^{\text{NL}}(t) = \varepsilon_0\left[\chi_{\text{el}}^{(3)} E_\phi(t)[E_\phi(t)]^2 + E_\phi(t) S_{\text{nu,a}}(t) + E_\phi(t) S_{\text{nu,b},\phi\phi}(t)\right], \tag{34}$$

$$P_\phi^{\text{NL}}(t) = \varepsilon_0 \chi_0^{(3)} E_\phi(t) \int_{-\infty}^{\infty} \left[f_{\text{el}}\,\delta(t-t') + f_{\text{nu,a}}\, g_{\text{nu,a}}(t-t') + f_{\text{nu,b}}\, g_{\text{nu,b}}(t-t')\right] E_\phi^2(t')\, dt', \tag{35}$$

which matches the simple example in [15], except this example is $\phi$-polarized instead of $x$-polarized. The Newton-Raphson method simplifies because of $\mathbf{E}^{n+1} = \hat{\boldsymbol{a}}_\phi\, E_\phi^{n+1}$, becoming the 1D Newton-Raphson method. The equations have the same form as the Cartesian TE$_z$ equations [5], after replacing $x$ with $r$ and $y$ with $\phi$. However, the $(\nabla \times \mathbf{E})_{z,\text{TE}}$ equation, used as a finite-difference update equation to determine $H_z$, is different from the Cartesian case.

### *2.6 TM$_z$ Equations*

For the special case of transverse to z magnetic fields, the nonlinear polarization vector $\mathbf{P}^{(3)}(t)$ simplifies, but it still includes anisotropic behavior,

$$P_r^{\text{NL}}(t) = \varepsilon_0\left[\chi_{\text{el}}^{(3)} E_r(t)|\mathbf{E}(t)|^2 + E_r(t) S_{\text{nu,a}}(\text{t}) + E_r(t) S_{\text{nu,b},rr}(t) + E_z(t) S_{\text{nu,b},rz}(t)\right], \tag{36}$$

$$P_z^{\text{NL}}(t) = \varepsilon_0\left[\chi_{el}^{(3)} E_z(t)|\mathbf{E}(t)|^2 + E_z(t) S_{\text{nu,a}}(\text{t}) + E_r(t) S_{\text{nu,b},rz}(t) + E_z(t) S_{\text{nu,b},zz}(t)\right]. \tag{37}$$

These equations show the tensor relationship between the electric field and polarization vector mentioned by Pohl in [7]. The Newton-Raphson method simplifies to $\mathbf{E}^{n+1} = \hat{\boldsymbol{a}}_r\, E_r^{n+1} + \hat{\boldsymbol{a}}_z\, E_z^{n+1}$, becoming the 2D Newton-Raphson method. The equations have the same form as the Cartesian TM$_z$ equations [5], replacing $x$ with $r$ and $y$ with $\phi$. However, the $(\nabla \times \mathbf{H})_{z,\text{TM}}$ equation, used in the Newton-Raphson method to determine $E_z$ and $E_r$, is different from the Cartesian case.

## 3. Numerical Simulation Parameters

This section shows examples of simulating TE$_z$ and TM$_z$ electromagnetic waves in the nonlinear isotropic materials BK7, fused silica and $CS_2$. First, a TE$_z$ simulation was run for BK7 and the results were compared with the coupled NLSE SSFM results from [17, 18]. Second, the example from [31] was extended adding the TE$_z$ simulation results for fused silica allowing for comparison with the prior TM$_z$ results, assuming the nonlinear polarization vector is isotropic. Third, simulations in fused silica using the "Combined TE$_z$ and TM$_z$ Equations" were performed for TE$_z$ and TM$_z$ sources "in phase" generating linear "transverse electric field vector polarization", sources $\pi/2$ "out of phase" generating circular "transverse electric field vector polarization", assuming the nonlinear polarization vector is isotropic. Note that the "transverse electric field vector polarization" is different than the "polarization vector" in the theory section. Lastly, $TE_z$ and $TM_z$ electromagnetic waves in $CS_2$ were simulated individually at approximately 1.28 MW peak input power, in a manner similar to Pohl's 1972 paper [7], appearing to confirm Pohl's 1972 paper [7] findings, that for narrower beams with smaller source spacings, $TM_z$ waves require more input power then comparable $TE_z$ waves to achieve similar results. The goal of each of these simulations was to show examples extending prior work [6-9], and to simulate electromagnetic waves propagating in a nonlinear isotropic material illustrating how the extended FDTD GVADE 3D CRS method is implemented.

The $TE_z$ and $TM_z$ fields were simulated using a *nr* x *nz* dimension Lebedev grid [32-34], in the same manner as [6]. Liu's paper [34] calls this grid the "unstaggered grid", which is a combination of two shifted Yee grids which results in collocated electric field components. The two Yee grids are coupled through the nonlinear polarization current terms. The simulation parameters $\Delta z$, $\Delta r$ and $\Delta t$ were chosen empirically by performing numerical convergence studies, attempting to balance efficiency and accuracy, with the resulting values chosen to be well below the Courant stability limit. The selection of these simulation parameters, the Courant stability limit and other simulation parameter related decisions are discussed in [5, 6] and described in further detail in [35].

### *3.1 Simulation Input Amplitude and Input Power*

For a meaningful comparison between the $TE_z$ and $TM_z$ simulations results, the input amplitude was adjusted so that the input power was the same for both $TE_z$ and $TM_z$, unless otherwise specified. For the $TM_z$ case, the input $H_\phi(z=0)$ amplitude $A_0$ required, to achieve the same input power as the $TE_z$ case, was determined using the plane wave approximation, $H_\phi(z=0) \approx E_\phi(z=0)/\eta$ where $\eta = \sqrt{\mu_0/(\varepsilon_0\varepsilon_r)}$ is the linear impedance of the material, where $\varepsilon_r = \sqrt{n_0}$ is the linear relative permittivity and $n_0$ is the linear refractive index determined using the Sellmeier expansion. Following this approach of determining the input amplitude, the $TE_z$ and $TM_z$ simulation results and behavior appeared to match for the same input power, except for the example in carbon disulfide. For the carbon disulfide example, for narrower beams using the same input power, differences between the $TE_z$ and $TM_z$ simulation results are clearly seen, requiring the $TM_z$ simulations to have a greater input power than the $TE_z$, to achieve similar simulation behavior. The plane wave approximation was used to approximate the input power, as seen in equations (38)-(44).

An input waveform used to excite the $TE_z$ electromagnetic wave at $z = 0$ is [6],

$$E_\phi(t, z=0) = V_{0,\mathrm{Sech}} \sin(\omega_c t + \beta_{\mathrm{TM}}) \left[\mathrm{sech}\left(\frac{r}{w_0} - spac\right) - \mathrm{sech}\left(\frac{r}{w_0} + spac\right)\right]. \quad (38)$$

Using the plane wave approximation to determine the approximate radial fields at z = 0:

$$\begin{aligned} &H_r(t, z=0) \\ &\approx \frac{V_{0,\mathrm{Sech}}}{\eta} \sin(\omega_c t + \beta_{\mathrm{TM}}) \left[\mathrm{sech}\left(\frac{r}{w_0} - spac\right) - \mathrm{sech}\left(\frac{r}{w_0} + spac\right)\right]. \end{aligned} \quad (39)$$

The instantaneous peak input power can then be approximated as,

$$\mathcal{P}_{z,\mathrm{TE}}^{\mathrm{Peak}} \approx 2\pi \frac{V_{0,\mathrm{Sech}}^{\;2}}{\eta} \int_{r=0}^{r=\infty} \left[\mathrm{sech}\left(\frac{r}{w_0} - spac\right) - \mathrm{sech}\left(\frac{r}{w_0} + spac\right)\right]^2 r\, dr. \quad (40)$$

The $\text{TM}_z$ equations for $H_\phi(t, z = 0)$, $\mathcal{P}_{z,\text{TM}}^{\text{Peak}}(t)$ and approximating $E_r(t, z = 0)$ are the duals of equations (38) to (40), with $E \rightarrow H$, $V \rightarrow A$, $1/\eta \rightarrow \eta$. The $\text{TE}_z$ and $\text{TM}_z$ sources can be $\Delta\beta = \beta_{\text{TE}} - \beta_{\text{TM}}$ radians "out of phase" of each other using $\sin(\omega_c t + \beta_{\text{TE}})$ for the $\text{TE}_z$ source.

As another input waveform, a vector Bessel-Gaussian azimuthally polarized beam can be approximated as a Laguerre-Gaussian amplitude profile at $z = 0$ [17, 18, 36]:

$$E_\phi(t, z = 0) = V_{0,\text{TEM}} \sin(\omega_c t)\, \rho e^{-\rho^2} \quad . \tag{41}$$

Using the plane wave approximation to determine the approximate radial fields at z = 0:

$$H_r(t, z = 0) = \frac{V_{0,\text{TEM}}}{\eta} \sin(\omega_c t)\, \rho e^{-\rho^2} \quad . \tag{42}$$

The instantaneous peak input power can then be approximated as,

$$\mathcal{P}_{z,\text{TE}}^{\text{Peak}} \approx 2\pi \frac{V_0{}^2}{\eta} \int_{r=0}^{r=\infty} [\rho e^{-\rho^2}]^2\, r\, dr\,, \tag{43}$$

where $\rho = 1$ corresponds to $1/e \approx 0.3679$ of the input amplitude $V_0$ and $\rho$ is defined as

$$\rho = r / r_0 \quad , \tag{44}$$

where $r_0$ is the "characteristic radius" or "spot radius".

The input waveforms using a $\rho e^{-\rho^2}$ radial distribution are shown in Fig. 1 for the $\text{TM}_z$ and $\text{TE}_z$ cases, which are radially polarized (RP) and azimuthally polarized (AP) respectively. The distribution $e^{-\rho^2}$ for the fundamental Gaussian mode [14, 37, 38] is shown in Fig. 1 for comparison purposes, since standard optics terms like the "diffraction length", the "characteristic radius" and the "angle of convergence" were defined for it, but they are also used as a standard of measure for other distributions like $\rho e^{-\rho^2}$, comparing them to the fundamental Gaussian mode [17, 37].

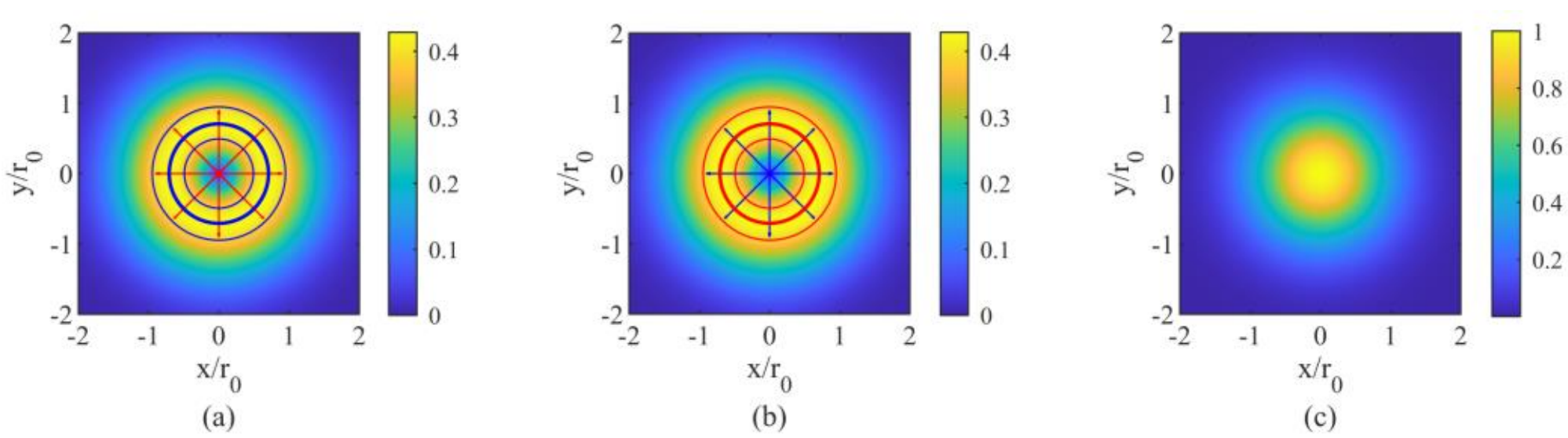

Fig. 1. The magnitude of the vector rotationally symmetric transverse fields are shown for $\text{TM}_z$ (RP) and $\text{TE}_z$ (AP) with the $\rho e^{-\rho^2}$ radial distribution at the input, $z = 0$, along with the magnitude of the fields for $e^{-\rho^2}$ non-vector rotationally symmetric radial distribution for comparison, where the solid red and solid blue curves added to the field plots are parallel to the electric and magnetic fields respectively. (a) The $\text{TM}_z$ input waveform $|H_\phi|$ is shown normalized to $V_0/\eta$. (b) The $\text{TE}_z$ input waveform $|E_\phi|$ is shown normalized to $V_0$. (c) The scalar $e^{-\rho^2}$ input distribution is shown.

Note that the time average power is related to the peak power above for an in-phase sinusoidal source by $\mathcal{P}_z^{\text{Peak}} = 2\mathcal{P}_z^{\text{Ave}}$.

### *3.2 Simulation and Material Parameters: BK7*

The isotropic polarization vector used by the 3D CRS FDTD GVADE method for BK7 is,

$$\mathbf{P}^{\mathrm{NL}}_{\mathrm{Isotropic}} = \varepsilon_0 \chi_0^{(3)} \mathbf{E}(\mathbf{r},t) \int_{-\infty}^{\infty} [\,\delta(t)(t-t')\,|\mathbf{E}(\mathbf{r},t')|^2] dt', \tag{45}$$

$$\mathbf{P}^{\mathrm{NL}}_{\mathrm{Isotropic}} = \varepsilon_0 \chi_0^{(3)} \mathbf{E}(\mathbf{r},t)|\mathbf{E}(\mathbf{r},t)|^2. \tag{46}$$

The input waveform parameters used were $V_{0,\mathrm{TEM}} = 2.2639\times10^{10}$ V and $r_0 = 4.8989\mu$m with a free space wavelength of $\lambda_0$=800nm, which corresponds to $\omega_c = 2.35456\times10^{15}$ rad/sec with a corresponding material linear wavelength of $\lambda_\mathrm{d} \approx 529.5$ nm (resulting in a 62:1 resolution, slightly exceeding the resolution from [4] using 8.5nm square grid cells). The linear index of refraction is determined using a 3 pole Sellmeier expansion with $\beta_1 = 1.03961212$, $\beta_2 = 0.231792344$, $\beta_3 = 1.01046945$, and $\omega_1 = 2.4316421\times10^{16}$ rad/sec, $\omega_2 = 1.3313467\times10^{16}$ rad/sec, $\omega_3$=$1.8509862\times10^{14}$ rad/sec [39]. Using the Sellmeier expansion, $\varepsilon_r(\omega_c) = [n_0(\omega_c)]^2 = 2.28244$. The nonlinear index of refraction for BK7 is modelled purely by the Kerr nonlinearity $n_{2,\mathrm{el}} = 3.3\times10^{-20}$ W/m$^2$ [40], which corresponds to an electronic third-order susceptibility of $\chi_0^{(3)} = \chi_{\mathrm{el}}^{(3)} = (4/3)c\,\varepsilon_0\varepsilon_r n_{2,\mathrm{el}}$.

The simulation parameters used were spatial grid steps of $\Delta z = \Delta r = 8.5$ nm for a $nr \times nz$ size simulation grid, where $nr = 3301$ and $nz = 25001$, with temporal steps of $\Delta t = 5.5\times10^{-18}$sec, for n = 100,000 time-steps. The time-step, $\Delta t$, was chosen empirically by performing numerical studies as 7.285 times below the required time-step for the Courant stability limit.

### *3.3 Simulation and Material Parameters: Fused Silica*

The isotropic polarization vector and response function used by the 3D CRS FDTD GVADE method for fused silica are,

$$\mathbf{P}^{\mathrm{NL}}_{\mathrm{Isotropic}} = \varepsilon_0 \chi_0^{(3)} \mathbf{E}(\mathbf{r},t) \int_{-\infty}^{\infty} \left[ g^{(3)}_{\mathrm{Isotropic}}(t-t')\,|\mathbf{E}(\mathbf{r},t')|^2 \right] dt', \tag{47}$$

$$g^{(3)}_{\mathrm{Isotropic}}(t) = \left(1 - f_{R,\mathrm{Isotropic}}\right)\delta(t) + f_{R,\mathrm{Isotropic}}\, g_a(t)\,. \tag{48}$$

For fused silica the material parameters used for this paper follow the model used in [1, 2, 6, 21] and are listed in [6]. The simulation parameters match [6], adding a TE input amplitude $V_0 = \eta A_0$, with $\eta = 256.82$, using the Sellmeier expansion $\varepsilon_r(\omega_c) = [n_0(\omega_c)]^2 = 2.1519$, for TM$_z$ and TE$_z$ with $n$ = 50,000 time-steps. The free space wavelength was selected as $\lambda_0$=433 nm to match [6], and the corresponding material linear wavelength of $\lambda_\mathrm{d} \approx 295.2$ nm (resulting in a 55.34:1 resolution following [4] using 5.333nm square grid cells). Please see Supplement 1 (Section 2: Fused Silica Rank 4 Susceptibility Tensor) for the rank 4 material response tensor written using index notation. The time-step, $\Delta t$, was chosen empirically by performing numerical studies as 7.527 times below the required time-step for the Courant stability limit.

### *3.4 Simulation and Material Parameters: Carbon Disulfide*

The polarization vector used by the 3D CRS FDTD GVADE method for carbon disulfide includes both the isotropic and anisotropic parts as described in [5]. The material nonlinear parameters used for this paper follow the model used in [41, 42], while the linear Sellmeier expansion parameters follow [43]. The material and simulation parameters match [5], with the exception that $\Delta r = \Delta x$, $nr = nx$ from [5], and $n$ = 50,000 time-steps. The simulation input/excitation parameters are listed in Table 1, where the input ampiltude $V_{0,\mathrm{Input},2000\mathrm{nm}}$ is either $V_{0,\text{Sech},2000\text{nm}}$=1.7967×10$^9$ V/m or $V_{0,\text{TEM},2000\text{nm}}$ = 3.9822×10$^9$ V/m with $w_0$= 2000nm and $r_0 = r_{0,3} = 2.4494 w_0$ respectively, as discussed in greater detail in Section 4.3. The vacuum carrier wavelength was selected at $\lambda_0$=694.3nm for a ruby laser, and the corresponding material linear wavelength of $\lambda_d \approx$ 430 nm (resulting in a 53.75:1 resolution following [4] using 8nm square grid cells), where $\varepsilon_r(\omega_c)$ =2.60488 and $\chi^{(3)}_{\text{el}} = (4/3)c\ \varepsilon_0\varepsilon_r n_{2,\text{el}}$ . The nuclear part of the susceptibility can be calculated using $\chi^{(3)}_{\text{nu}}(t) = 2\ c\ \varepsilon_0\varepsilon_r\ R(t)$ where $R(t) = \sum_m n_{2,m} g_m(t)$ as described in [41, 44]. To determine $\chi^{(3)}_0$, $f_{\text{el}}$, $f_c$, $f_l$ and $f_d$, the special case where $i = j = k = l = x$ was considered leading to,

$$\chi^{(3)}_{xxxx}(t) = (4/3)c\ \varepsilon_0\varepsilon_r n_{2,el} g_{el}(t) + 2\ c\ \varepsilon_0\varepsilon_r\big[n_{2,c} g_c(t) + n_{2,l} g_l(t) + n_{2,d} g_d(t)\big]. \tag{49}$$

The equation can be re-written pulling out a constant $\chi^{(3)}_0$ as $\chi^{(3)}_{xxxx}(t) = \chi^{(3)}_0 g_{xxxx}(t)$,

$$\chi^{(3)}_{xxxx}(t) = \chi^{(3)}_0 [f_{el} g_{el}(t) + f_c g_c(t) + f_l g_l(t) + f_d g_d(t)]\ , \tag{50}$$

where $f_{el} = (2/3)n_{2,\text{el}}/n_{2,\text{sum}}$, $f_c = n_{2,c}/n_{2,\text{sum}}$, $f_l = n_{2,l}/n_{2,\text{sum}}$, $f_d = n_{2,d}/n_{2,\text{sum}}$, $\chi^{(3)}_0 = 2\ c\ \varepsilon_0\varepsilon_r n_{2,sum}$, and $n_{2,\text{sum}} = (2/3)n_{2,el} + n_{2,c} + n_{2,l} + n_{2,d}$ . The resulting values were $\chi^{(3)}_0$ = 3.81677×10$^{-20}$m$^2$/V$^2$, $f_{\text{el}}$ = 0.0362, $f_c$= 0.0362, $f_l$ = 0.2754 and $f_d$ = 0.65217. Please see Supplement 1 (Section 3: Carbon Disulfide Rank 4 Susceptibility Tensor) for the rank 4 material response tensor written using index notation and for more details on how $\chi^{(3)}_0$, $f_{el}$, $f_c$, $f_l$ and $f_d$ were determined. The time-step, $\Delta t$, was chosen empirically by performing numerical studies as 7.051 times below the required time-step for the Courant stability limit.

## 4. Simulation Results and Discussion

This section shows 3D CRS FDTD GVADE simulation results of rotationally symmetric wave propagation in three different nonlinear materials: isotropic BK7 which has only Kerr nonlinearity; isotropic fused silica which has both Kerr and Raman nonlinearities; and $CS_2$ which has 4 nonlinearities, where two nonlinearities are isotropic and two nonlinearities are anisotropic. The first example in BK7 compares "finite time" 3D CRS FDTD GVADE simulation results with "steady-state" NLSE SSFM simulation results [17, 18, 45]. The next two examples compare TE$_z$ and TM$_z$ 3D CRS FDTD GVADE simulation results in fused silica, comparing with prior work from [6], and simulating the "Combined TE$_z$ and TM$_z$" fields as described in Section 2.4. The last example compares TE$_z$ and TM$_z$ 3D CRS FDTD GVADE simulation results in carbon disulfide with Pohl's experimental results [7].

### *4.1 Simulation Results and Discussion – BK7*

The first example simulation is of a TE$_z$ source in BK7, as shown in Eq. (41), which has azimuthal polarization (AP). The simulation results obtained using the 3D CRS FDTD GVADE method are compared with the SSFM NLSE Matlab simulation results [17, 18]; the Matlab code was translated from the Fortran code generously provided by Dr. Amiel A. Ishaaya [45]. In [17, 18], their Fig. 1a simulates a TE$_z$ wave with a time average input power of $\mathcal{P}_z^{\text{Ave}} = 10\mathcal{P}_z^{\text{Critical Power Gaussian}}$ with a sinusoidal source, which corresponds to an instantaneous peak power of $\mathcal{P}_z^{\text{Peak}} = 2\mathcal{P}_z^{\text{Ave}} = 2\left(10\mathcal{P}_z^{\text{Critical Power Gaussian}}\right) = 20\mathcal{P}_z^{\text{Critical Power Gaussian}}$ [14]. The simulation results in Fig. 2 compare the wave propagation simulated using both methods. Fig. 2a shows the 3D CRS FDTD GVADE simulated |E|, while Fig. 2b compares the radial extremum locations of the waves as they propagate, in the "initial and intermediate collapse regimes" [17] not showing the extremum locations beyond that point. An important difference between the 3D CRS FDTD GVADE method and the SSFM NLSE results is that FDTD based method simulation results are at a set time, $n\Delta t$, while the NLSE equations are generally at "steady state" assuming the wave has been propagating long enough for "steady state" to be reached.

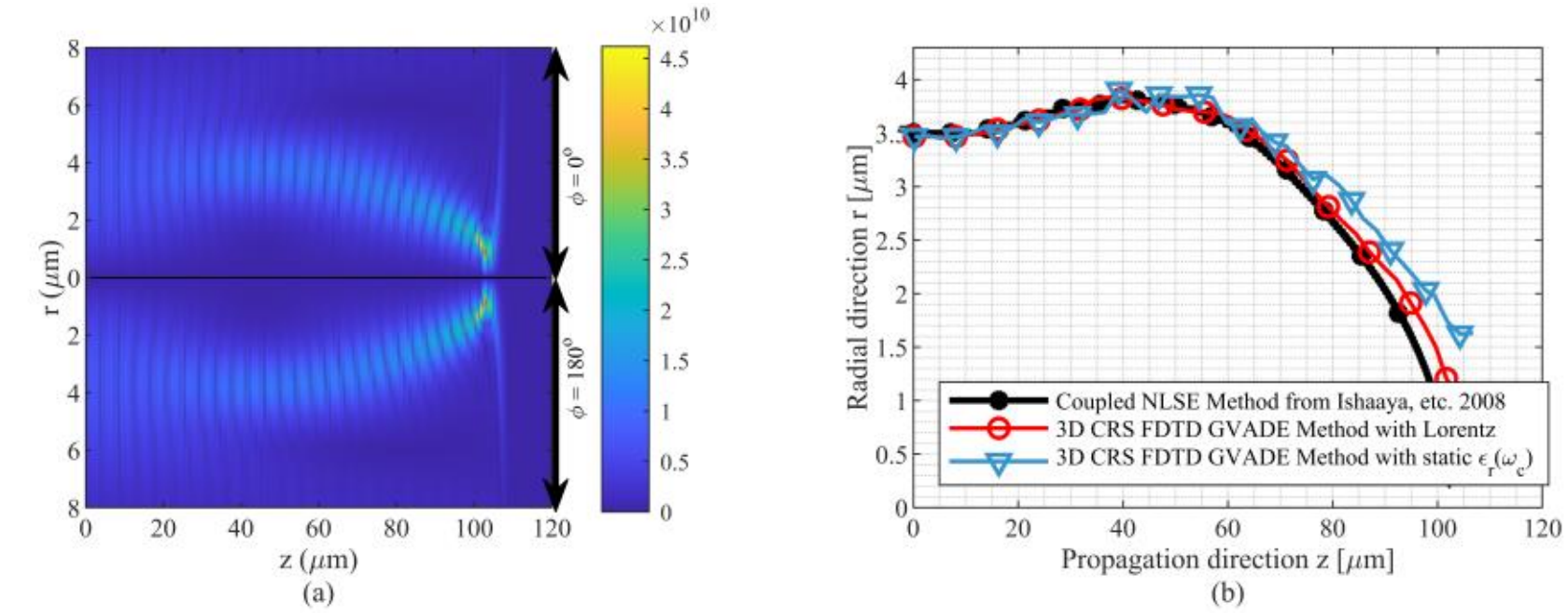

Fig. 2. Comparing the simulation results of the 3D CRS FDTD GVADE method and the SSFM NLSE method [17, 18, 45] in the $y$ = 0 plane for a $TE_z$ source/excitation of the form $E_\phi(t, z = 0) = V_{0,\mathrm{TEM}}\rho e^{-\rho^2}$. (a) |**E**| for a $TE_z$ wave in BK7 using the 3D CRS FDTD GVADE method, assuming a purely isotropic polarization vector with n = 100,000 time-steps, $\Delta t = 5.5\times10^{-18}$sec and $\Delta r = \Delta z = 8.5\times10^{-9}$meters, including the linear Lorentz dispersion, (b) Soliton extremum locations of $E_\phi$ in the $r$ direction for the 3D CRS FDTD GVADE method, both with constant/static $\varepsilon_r(\omega_c)$ and with Lorentz dispersion, compared with the extremum locations of $E_\phi$ in the $r$ direction of the SSFM NLSE method results versus propagation direction $z$. Note: Due to the input power being significantly larger than the critical power, the 3D CRS FDTD GVADE's Newton-Raphson method may experience double precision computing limitations near $r = 0$ if the simulation continued indefinitely as will be discussed further in [35].

By comparing the simulations results in Fig. 2b, there is a close agreement between the 3D CRS FDTD GVADE extremum locations and the NLSE SSFM extremum locations. The same trend is seen in the wave propagation for both numerical methods, with both methods showing the same initially expanding and then collapsing behavior. This example was a good comparison with the NLSE SSFM due to the presence of the most basic/simple nonlinearity, the isotropic "instantaneous" electronic response, which seems to be modeled well by the NLSE SSFM. Next, the following sections will explore simulations of material with more complicated nonlinear responses.

### *4.2 Simulation Results and Discussion – Fused Silica*

The second example results, simulating the $TE_z$ waves in silica, for comparison with the $TM_z$ results from [6], are shown in Fig. 3. The extremum locations and the smoothed transverse electric field amplitudes are compared for $TE_z$ and $TM_z$. By observation, for this example, the $TE_z$ and $TM_z$ simulation results are remarkably similar, appearing visually indistinguishable.

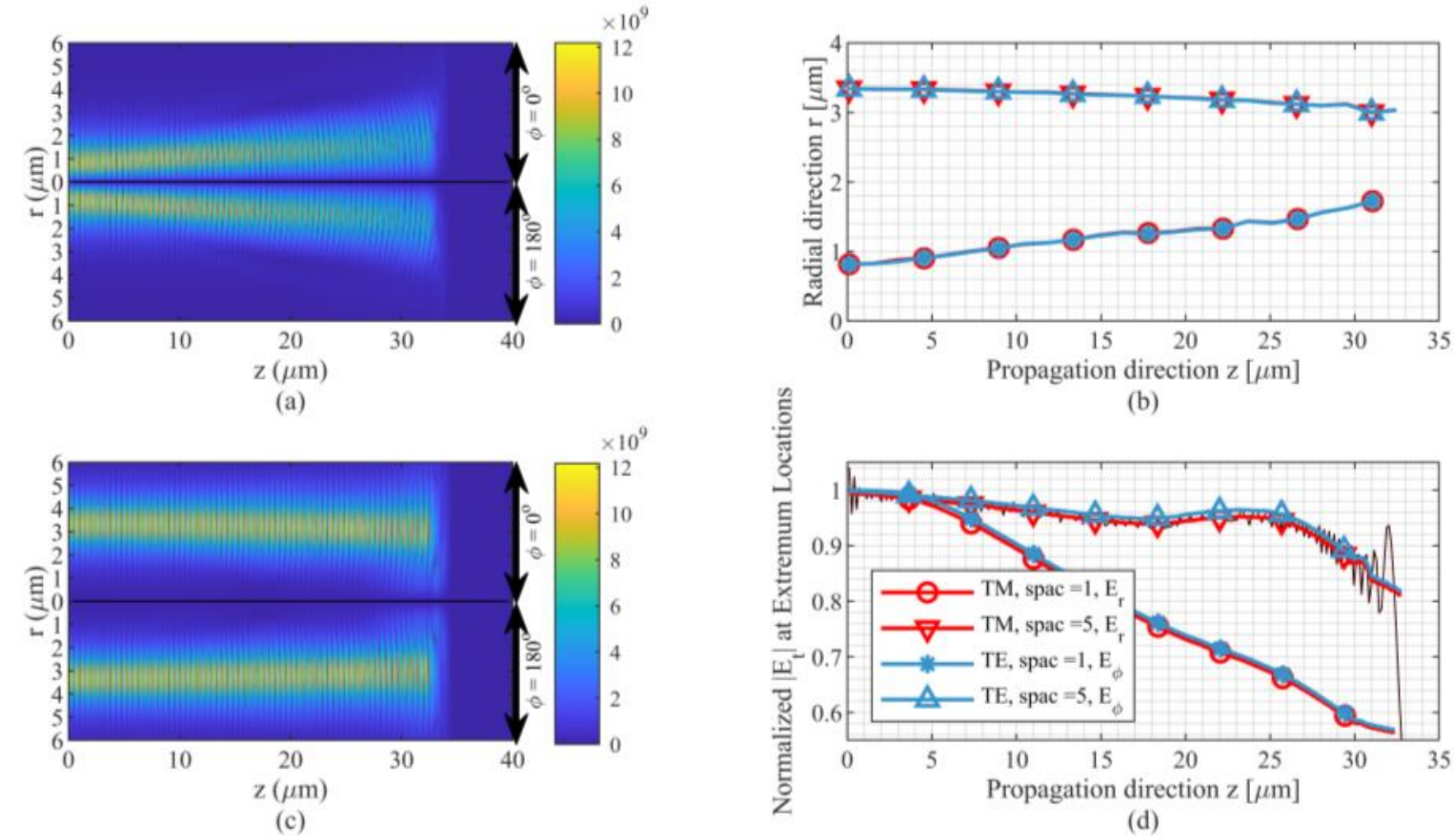

Fig. 3. Comparing the simulation results of $|\mathbf{E}|$ for $\mathrm{TE}_z$ and $\mathrm{TM}_z$ in fused silica for source spacings of 1 and 5 with n = 50,000 time-steps and $\Delta t = 3.34\times10^{-18}$sec and $\Delta r = \Delta z = 16/3\times10^{-9}$sec. (a) $\mathrm{TE}_z$ with source spacing = 1, (c) $\mathrm{TE}_z$ with source spacing = 5, (b) Soliton extremum locations from $\sin(\omega_c t)$ of $E_\phi$ and $H_\phi$ in the $r$ direction versus propagation direction $z$ in the $y$ = 0 plane for $\mathrm{TE}_z$ and $\mathrm{TM}_z$ with source spacing = 1 and 5, (d) Smoothed transverse electric field magnitudes $|E_\phi|$ and $|E_r|$ at the extremum locations from $\sin(\omega_c t)$ of $E_\phi$ and $H_\phi$ versus propagation direction $z$ in the $y$ = 0 plane for $\mathrm{TE}_z$ and $\mathrm{TM}_z$ with source spacing = 1 and 5, normalized to $|E_\phi(z = z_{E_\phi\, extremum\, nearest\, \mathrm{z=0}})|$. The thinner darker red curve is the "unsmoothed" $|E_r|$ at the extremum locations of $H_\phi$ curve for source spacing = 5 as an example of the moving-mean smoothing process being used. Note 1: The $\mathrm{TM}_z$ simulation results of $|\mathbf{E}|$ corresponding to (a) and (c) can be found in [6]. Note 2: Due to the large extremum location density, the plot markers in (b) are at evenly sampled locations, while maintaining the shape of the curves.

The third example, also performed in fused silica, simulated "Combined $\mathrm{TE}_z$ and $\mathrm{TM}_z$" waves, assuming an isotropic polarization vector, as shown in Fig. 4 and Fig. 5 with a phase difference of $\Delta\beta = 0$ and a phase difference of $\Delta\beta = \pi/2$ between the $\mathrm{TE}_z$ and the $\mathrm{TM}_z$ sources.

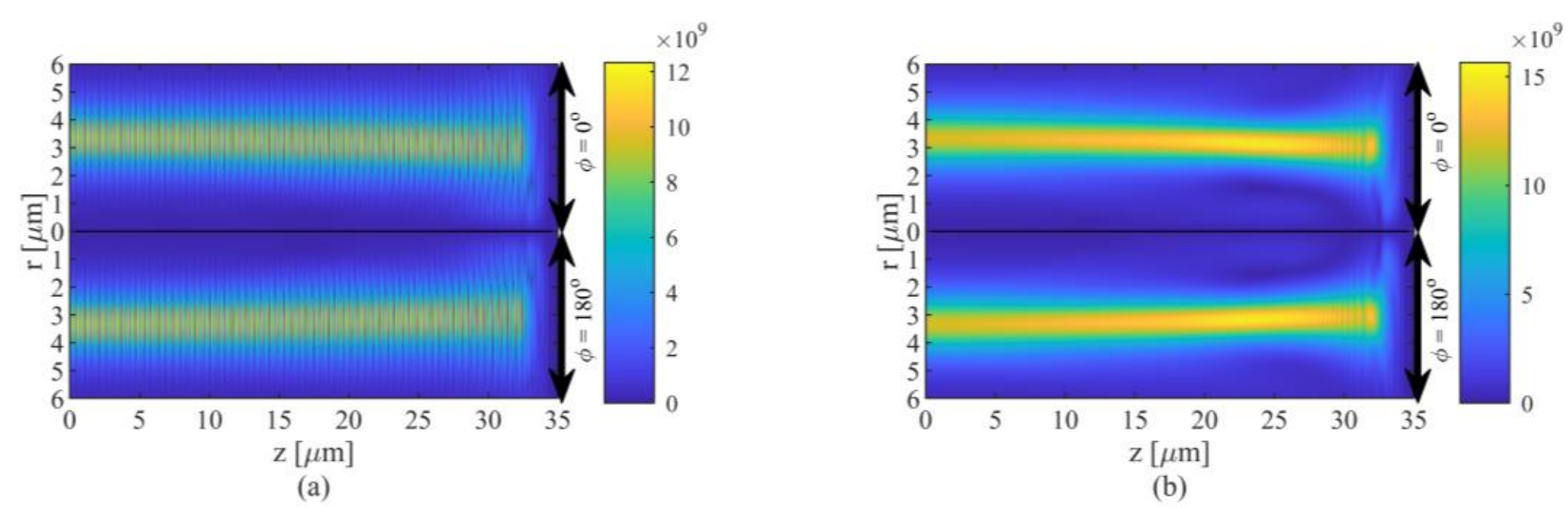

Fig. 4. Comparing the simulation results of $|\mathbf{E}|$ for "Combined $\mathrm{TE}_z$ and $\mathrm{TM}_z$" waves in fused silica for $\mathrm{TE}_z$ and $\mathrm{TM}_z$ sources with $\Delta\beta = 0$ and $\Delta\beta = \pi/2$ phase differences and a source spacing = 5, assuming a purely isotropic polarization vector with n = 50,000 time-steps, $\Delta t = 3.34\times10^{-18}$sec and $\Delta \mathrm{r} = \Delta \mathrm{z} = 16/3\times10^{-9}$meters. (a) Combined $\mathrm{TE}_z$ and $\mathrm{TM}_z$ with $A_0/\sqrt{2}$ and $V_0/\sqrt{2}$ using $\Delta\beta = 0$, leading to linear polarization, (b) Combined $\mathrm{TE}_z$ and $\mathrm{TM}_z$ with $A_0$ and $V_0$ using $\Delta\beta = \pi/2$, leading to approximately circular polarization.

The simulation results are shown in Fig. 5a and Fig. 5c from z = 26μm to z = 30μm to allow the individual extremums of $|\mathbf{E}|$ from $\sin(\omega_c t)$ to be seen clearly in Fig. 5a where $\Delta\beta = 0$.

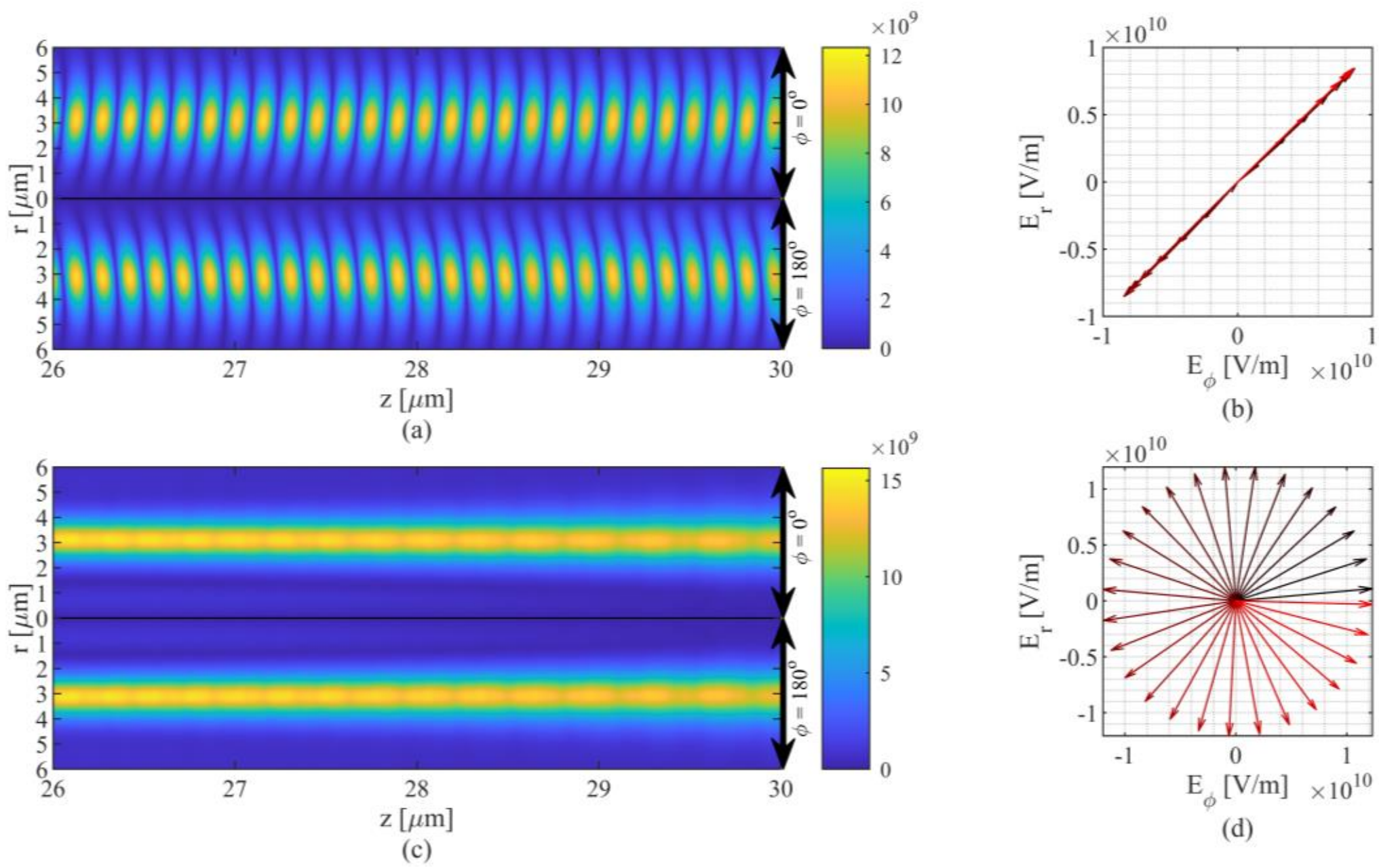

Fig. 5. Comparing the simulation results of $|\mathbf{E}|$ for "Combined $TE_z$ and $TM_z$" waves in silica for $TE_z$ and $TM_z$ sources with $\Delta\beta = 0$ and $\Delta\beta = \pi/2$ phase differences and a source spacing = 5, assuming a purely isotropic polarization vector with n = 50,000 time-steps, $\Delta t = 3.34\times10^{-18}$sec and $\Delta r = \Delta z = 16/3\times10^{-9}$meters. (a-b) Combined $TE_z$ and $TM_z$ with $A_0/\sqrt{2}$ and $V_0/\sqrt{2}$ using $\Delta\beta = 0$, leading to linear polarization, (c-d) Combined $TE_z$ and $TM_z$ with $A_0$ and $V_0$ using $\Delta\beta = \pi/2$, leading to approximately circular polarization. The electric field vector polarization with wave propagation in space of the "Combined $TE_z$ and $TM_z$" is seen for both simulations in (b) and (d) by looking at the combined transverse electric field vector $\mathbf{E}_t = \hat{\boldsymbol{a}}_r E_r + \hat{\boldsymbol{a}}_\phi E_\phi$ at $r_{\text{extremum}} \approx 3.344$ μm, for sampled $z$ grid locations spanning one wavelength $\lambda_d = \lambda_0/\sqrt{\varepsilon_r}$, where "pure black" corresponds to $z = z_{\text{extremum location,5}}$ and "pure red" corresponds to $z_{\text{extremum location,7}} = (z_{\text{extremum location,5}} + \lambda_d)$, at a fixed time $t$ = (50,000 time-steps)$\Delta t$.

For the $\Delta\beta = 0$ simulation, Fig. 5b shows the polarization is linear. For the $\Delta\beta = \pi/2$ simulation, Fig. 5d shows the polarization to be approximately circular, due to $E_t \approx E_r \approx E_\phi$. Note that $|\mathbf{E}|$ in Fig. 5c does not allow the individual extremums of the $TE_z$ and the $TM_z$ waves to be clearly seen, since $|\mathbf{E}_t|$ is approximately constant at $r_{\text{extremum}}$, $|\mathbf{E}_t| \approx E_t \sqrt{\sin^2(\omega_c t) + \cos^2(\omega_c t)} = E_t$ .

### *4.3 Simulation Results and Discussion – Carbon Disulfide*

The last example uses Pohl's experimental measurements from [7] for validation, simulating the $TE_z$ and $TM_z$ waves individually in $CS_2$ for comparison, as shown in Fig. 6 and Fig. 7a, for both input waveforms in Eq. (38) and Eq. (41). Comparing the simulation results for the two different input waveforms was accomplished by adjusting the input amplitudes to achieve the same input power, 1.28MW "peak-power" using equations (40) and (43), along with choosing a common input waveform radius or "spot size". There were two possible choices for how to define the input waveform radius based on the usage in [7] and its references [12, 46, 47], as the radial distance required for the input waveform amplitude to drop by a $1/e$ factor from the amplitude maximum located at $r = 0$ for the fundamental $TEM_{00}$ Gaussian beam. The two possible radius definitions using the assumed $\rho e^{-\rho^2}$ field distribution of [7] instead of the $TEM_{00}$ are: 1) the radial distance where the azimuthal field dropped to $0.85776 \max[E_\phi(z = 0)]$ or $0.85776\,\max[H_\phi(z=0)]$, which corresponds to $1/e$ for $\rho e^{-\rho^2}$ at $\rho = 1$ as shown in Fig. 6d as $r_{0,3}$ and $r_{0,2}$, and 2) the radial distance where the azimuthal field dropped to $1/e$ of its maximum, $\max[E_\phi(z=0)]/e$ or $\max[H_\phi(z=0)]/e$ as shown in Fig. 6d as $r_{0,1}$; the maximum value of $\rho e^{-\rho^2}$ is approximately 0.42888 rather than 1 and the maximum value of $[\mathrm{sech}(r/w_0-2)-\mathrm{sech}(r/w_0+2)]$ is approximately 0.9639 rather than 1, meaning that $1/e$ at $\rho = 1$ for $\rho e^{-\rho^2}$ corresponds to 0.85776 of $0.42888V_0$ and $1/e$ for $[\mathrm{sech}(r/w_0-2)-\mathrm{sech}(r/w_0+2)]$ corresponds to 0.381657 of $0.9639V_0$, rather than $1/e$ of 1. The first definition of the radius or "spot size" appears more likely from the mathematical definitions in [7], while the second definition may make more intuitive sense when considering the term "spot size" while observing Fig. 6d from an experimental perspective, measuring the $1/e$ factor relative to the maximum rather than the $TEM_{00}$ theoretical maximum, but it was not clear to the authors how "spot size" and angle of convergence were determined experimentally in [7]. As a result, both of the above definitions for the radius or "spot size" are used in Fig. 7a for comparison. For Fig. 6, the second definition was used, while also illustrating the second definition in Fig. 6d, where both of the input waveforms were set to have the same radius $r_{0,1} = 3.6763w_0$ at $\max(E_\phi)/e$ or $\max[H_\phi(z=0)]/e$, meaning $r_{0,2} \neq r_{0,3}$, with $r_{0,3} = 2.44945w_0$. The other option would have been to set $r_{0,2} = r_{0,3}$, requiring one of the input waveform amplitudes to have been adjusted to maintain the same input power.

For the specific case where $w_0$=1000nm and $r_{0,1} = 3.6763$µm, the simulation results are shown in Fig. 6b, and the $TE_z$ and $TM_z$ simulation results are remarkably similar again, with the radial direction extremum locations visually appearing to be identical in Fig. 6b, with a "root-mean-square relative error" (RMSRE) of less than 0.006 as shown in Table 1 for both input waveforms. Also by observation of Fig. 6b, the propagation behavior of the two input waveforms is similar. The results shown in Fig. 6 for the $\rho e^{-\rho^2}$ and the $[\mathrm{sech}(r/w_0-2)-\mathrm{sech}(r/w_0+2)]$ input waveforms each represent two data points in Fig. 7a respectively, at $r_{0,i}$ and $r_{0,i,min}$, as defined below. However, for smaller $w_0$ and $r_{0,1}$ values, the $TE_z$ and $TM_z$ simulation results can be observed to diverge when the $TE_z$ and $TM_z$ input power are the same as shown in Fig. 7b.

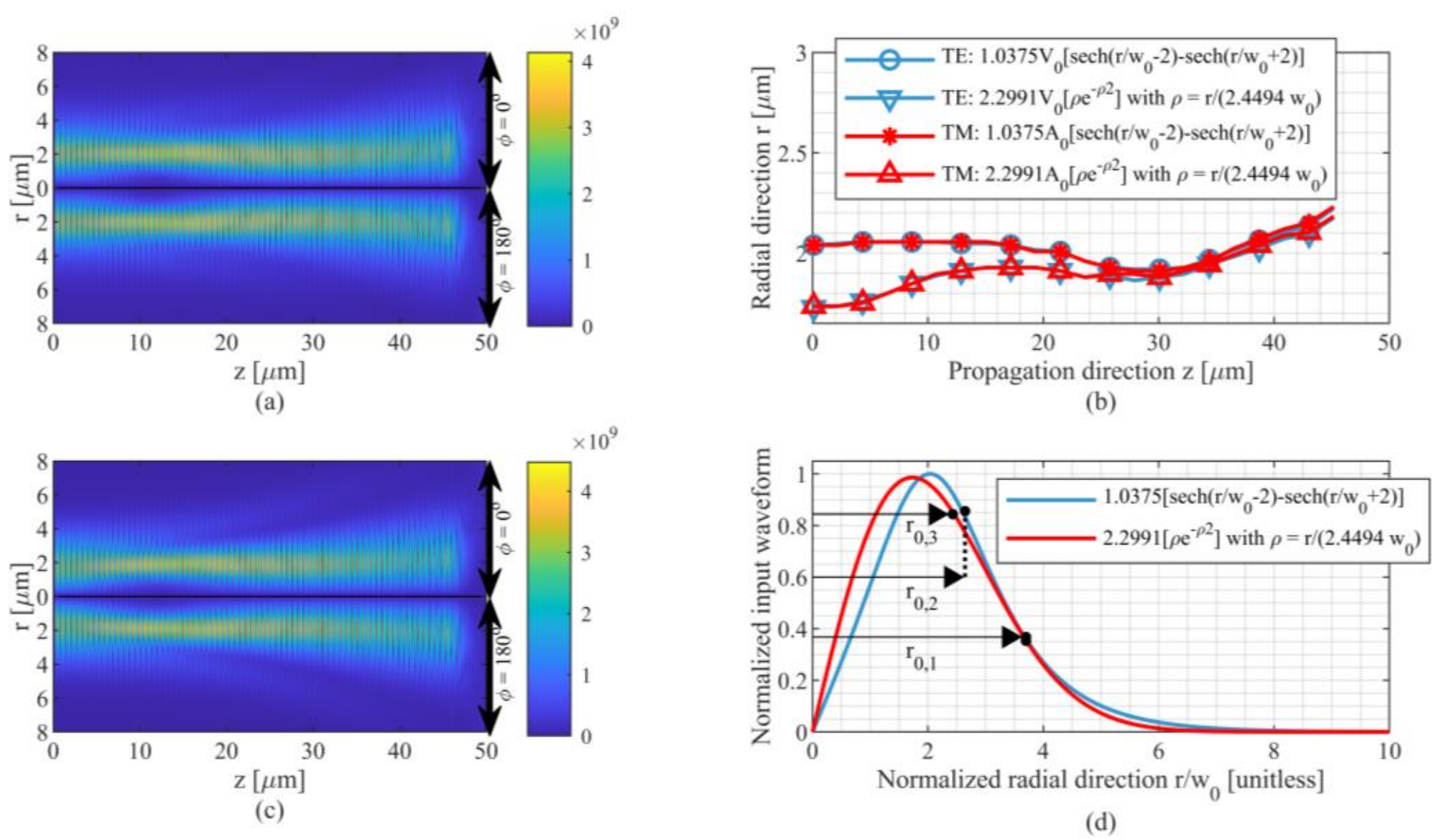

Fig. 6. Comparing the simulation results of $TE_z$ and $TM_z$ for both types of input waveforms with $w_0$=1000nm and $r_{0,1} = 3.6763w_0 = 3.6763\mu$m with $n$ = 50,000 time-steps, $\Delta t = 5.3486\times10^{-18}$sec and $\Delta r = \Delta z = 8\times10^{-9}$m. (a) $|\mathbf{E}|$ for $TE_z$ simulation with $1.0375V_0[\text{sech}(r/w_0 - 2) - \text{sech}(r/w_0 + 2)]$ input waveform and source spacing = 2, (c) $|\mathbf{E}|$ for $TE_z$ simulation with $2.2991V_0[\rho e^{-\rho^2}]$ input waveform and $r_{0,3} = 2.4494w_0$, (b) Soliton extremum locations from $\sin(\omega_c t)$ of $E_\phi$ and $H_\phi$ in the $r$ direction versus propagation direction $z$ in the $y = 0$ plane for $TE_z$ and $TM_z$ for both input waveforms, (d) Azimuthal electric field, $E_\phi$, normalized to $V_0$, at the source location z = 0. Note: Due to the large extremum location density, the plot markers in (b) are at evenly sampled locations, while maintaining the shape of the curves.

By varying $r_{0,1}$ as described in Table 1, this example in $CS_2$ attempts to compare Pohl's experimentally measured data points in Fig. 1 from [7] with simulated results using our 3D CRS FDTD GVADE method as shown in Fig. 7a. Note that significantly shorter numerically simulated 267.43 fsec pulses were compared with Pohl's experimentally measured "1-nesec pulses" data points for two reasons: first, due to the greater computational resources required by FDTD based numerical method, there are known practical limitations on the pulse duration which can be feasibly simulated; second, to the knowledge of the authors, other than Pohl's 1972 paper, no other known vector rotationally symmetric experimental data of wave propagation in nonlinear optically anisotropic material was available to compare with for validation. Fig. 7a compares the input power required for $TE_z$ and $TM_z$ waves propagating in $CS_2$ to achieve minimal error between their radial direction extremum locations as shown in Table 1. In Fig. 7a the ratio of the $TM_z$ and $TE_z$ input power is shown for two different radius definitions, $r_{0,1}$ and $r_{0,2}$ ( $r_{0,3}$) at $z = 0$, for the$[\text{sech}(r/w_0 - 2) - \text{sech}(r/w_0 + 2)]$ ($\rho e^{-\rho^2}$) input waveform, with the corresponding error between the $TM_z$ and $TE_z$ radial extremum locations shown in Table 1 for both of the input waveforms. The ratio of the $TM_z$ and $TE_z$ input power was also plotted at the z location, of the extremum location minimum, $z = z_{r_{0,\text{i,min}}}$, using the radius at that z location, $r_{0,\text{i,min}}$, as shown in Fig. 7b, for both of the input waveforms. The authors were uncertain which of the curves shown in Fig. 7a would best match Pohl's experimental configuration, and as a result all four were included.

**Table 1. TE and TM Input Amplitudes and Extremum Locations Comparison, Input = "TEM" or "Sech"**

| $w_0$ /(2000nm) | $r_{0,1}$ ($\mu$m) | $V_{0,\text{Input}}$ / ($V_{0,\text{Input,2000nm}}$) | $A_{0,\text{Input}}$ / $\left(\frac{V_{0,\text{Input}}}{\eta}\right)$ | $P^{\text{Peak}}_{\text{Input,TM}}$ /(1.28MW) | RMSRE Sech | RMSRE LG |
|---|---|---|---|---|---|---|
| 1 | 7.36 | 1 | 1 | 1 | 0.0014 | |
| 1/2 | 3.68 | 2 | 1 | 1 | 0.0058 | 0.0053 |
| 1/4 | 1.84 | 4 | 1.05 | 1.1025 | 0.0133 | 0.0136 |
| 1/6 | 1.23 | 6 | 1.1 | 1.21 | 0.0175 | |
| 1/8 | 0.92 | 8 | 1.15 | 1.3225 | 0.0453 | 0.0612 |
| 1/10 | 0.736 | 10 | 1.2 | 1.44 | 0.0672 | |

In Fig. 7a, measured $P_{\text{cr,TM}}/P_{\text{cr,TE}}$ from [7] are shown, where $P_{\text{cr}}$ is the "critical power" which was determined experimentally by Pohl, and it is stated in [7] that $P_{\text{cr,TM}}$ and $P_{\text{cr,TE}}$ are greater than " 1-MW peak". Determining $P_{\text{cr,TM}}$ and $P_{\text{cr,TE}}$ numerically is more difficult than experimentally. So for our numerical method, due to having finite computational resources, the ratio of $P_{\text{TM}}/P_{\text{TE}}$ is plotted with a constant $P_{\text{TE}}$ = 1.28MW, and it was assumed that the ratio of the critical power's shown in Pohl's paper would be a reasonable comparison to the ratio of the input powers with roughly equivalent radial extremum locations. The "angle of convergence" used in Fig. 7a is defined as $\theta = r_0 / L_{\text{diff}}$ and it is the small angle approximation for $\tan(\theta) = r_0 / L_{\text{diff}}$ used in the paraxial approximation with Gaussian beams, with $L_{\text{diff}} = (k\, r_0^2)/2$ being the diffraction length, $k = 2\pi/\lambda$ being the wavenumber and $\lambda$ being the wavelength in the material. An example of $TE_z$ and $TM_z$ waves requiring greater $TM_z$ input power to achieve minimal error in their radial extremum locations compared to the $TE_z$ is shown in Fig. 7b for the $r_{0,1} = 0.92$ nm for the $[\text{sech}(r/w_0 - 2) - \text{sech}(r/w_0 + 2)]$ input waveform with an RMSRE of 0.1456 for $P_{\text{TM}} = P_{\text{TE}}$, and an RMSRE of 0.0453 for the better curve fit where $P_{\text{TM}} = 1.32P_{\text{TE}}$. The last difference that should be noted between Pohl's experimental results and our numerical method is that our numerical method does not model plasma formation, whereas Pohl describes plasma as appearing for $\theta > 0.10$, which may account for the drop in $P_{\text{TM}}/P_{\text{TE}}$ seen in Pohl's data in Fig. 7a from $\theta \approx 0.15$ to $\theta \approx 0.2$.

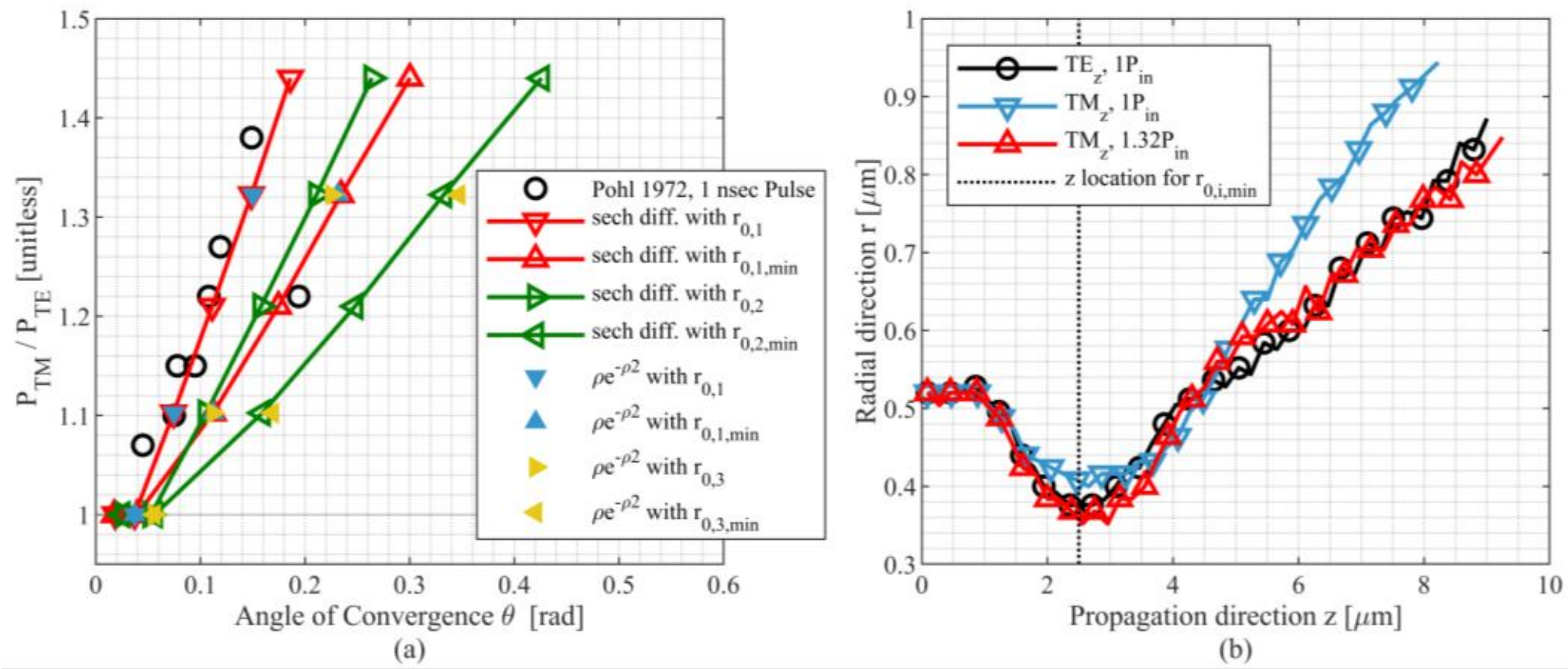

Fig. 7. Comparing the simulation results of $TE_z$ and $TM_z$ input powers required to achieve similar extremum locations with a constant $P_{\text{TE}} = 1.28$MW with $n = 50{,}000$ time-steps, $\Delta t = 5.3486\times10^{-18}$sec and $\Delta r = \Delta z = 8\times10^{-9}$m. (a) $P_{\text{TM}}/P_{\text{TE}}$ versus the "Angle of Convergence" with measured "1-nsec pulses" data points taken from Fig. 1 in [7], with $r_{0,2} = r_{0,1}/1.418312$ and $r_{0,3} = r_{0,1}/1.500875$ (b) The extremum locations of $E_\phi$ and $H_\phi$ waves in the $r$ direction versus propagation direction $z$ for $TE_z$ and $TM_z$ with the $1.0375[\text{sech}(r/w_0 - 2) - \text{sech}(r/w_0 + 2)]$ input waveform, with input amplitudes $V_0$, $A_0$ and $1.15A_0$ corresponding to $1P_{\text{in}}$, $1P_{\text{in}}$ and $(1.15)^2P_{\text{in}}$ where $(1.15)^2 = 1.3225$, with $w_0$=250nm and $r_{0,1} = 0.92$ nm and $r_{0,\text{i,min}}$ corresponding to $r_{0,1,\text{min}}$ and $r_{0,2,\text{min}}$ respectively.

It should be noted that for each data point in Fig. 7a, both a $TE_z$ and $TM_z$ simulation were required to be run. The $TE_z$ simulation on average took roughly 3-4 days to complete while the $TM_z$ simulation required closer to 7 days, on the Texas A&M Grace supercomputer, an Intel x86-64 Linux cluster. In this case, a simulation day roughly corresponds to 1,100 CPU hours. The longer simulation time of the $TM_z$ case is due to the fact that there are two electric field components along with their corresponding auxiliary variables, meaning that there were more arrays required corresponding to larger memory requirements, slowing the simulation times, rather than the one electric field component required for $TE_z$ with its fewer corresponding auxiliary variables and arrays. Also, the $TM_z$ simulations use a 2D Newton-Raphson method while the $TE_z$ simulations only required the 1D Newton-Raphson method to determine the electric field at the new time-step, further slowing $TM_z$ simulation in comparison to the $TE_z$.

The 3D CRS FDTD GVADE method simulation results shown in Fig. 7a appear to follow the same trend of Pohl's experimental measurements, thus validating the numerical method. The difference in pulse duration between our simulation and Pohl's experimental measurements, along with the 3D CRS FDTD GVADE method not modeling plasma, seem reasonable explanations for any differences between the experimental simulated results. By observation of Fig. 7a, all the curves seem to follow the same general trend as Pohl's experimental results, while the effective slope of the curves vary with the choice of the definition of radius.

## 5. Conclusion

In this paper we presented simulations of 3D full wave time domain electromagnetic spatial solitons propagating in a nonlinear dispersive isotropic media with azimuthal and radial transverse electric fields. The method for construction and analysis of $TE_z$ and $TM_z$ rotationally symmetric 3D electromagnetic waves, along with the combined $TE_z$ and $TM_z$ rotationally symmetric 3D electromagnetic wave method was developed using a two-dimensional numerical grid with the FDTD GVADE method modified for cylindrical rotationally symmetric problems. The simulations in fused silica extended prior work, adding the $TE_z$ rotationally symmetric 3D electromagnetic wave to the $TM_z$, and simulated linear electric field vector polarization and approximately circular electric field vector polarization using the combined $TE_z$ and $TM_z$ rotationally symmetric 3D electromagnetic wave. In BK7, the 3D CRS FDTD GVADE method simulation results were compared with the NLSE SSFM simulations results, showing comparable wave propagation behavior, validating the numerical method for isotropic nonlinear Kerr electronic response. In carbon disulfide, similar results to Dieter Pohl's original experimental and theoretical work on vectorial rotationally symmetric $TE_z$ and $TM_z$ self-trapping light beams from the early 1970s were numerically produced, validating the 3D CRS FDTD GVADE method, extending the polarization vector to include an anisotropic part for the transverse magnetic case, where the electric field vector and polarization vector are tensorially related.

**Acknowledgments.** Special thanks to Dr. Amiel A. Ishaaya for his email correspondence, and for sending his Fortran code from his 2008 OPTICS LETTERS paper "Self-focusing dynamics of polarization vortices in Kerr media" [17, 18, 45] helping to further validate the numerical method. Caleb Grimms thanks Junseob Kim for his contribution by creating the initial 2D Cartesian isotropic polarization vector FDTD GVADE code for his master's thesis based on [1, 2, 4].

**Disclosures.** The authors declare no conflicts of interest.

**Data availability.** To the best of the authors' knowledge, the data underlying the simulation results and figures presented in this paper are included in this paper along with [5, 6, 35], but more information may also be obtained from the authors upon reasonable request. The Fortran code used to compare with the method presented in this paper is the property of A. A. Ishaaya [17, 45] and should be requested directly from him, and as a result, the authors' translated Matlab code is not available.

**Supplemental document.** See Supplement 1 for supporting content.

## GENERAL VECTOR AUXILIARY DIFFERENTIAL EQUATION FINITE-DIFFERENCE TIME-DOMAIN METHOD FOR ROTATIONALLY SYMMETRIC VECTOR WAVE PROPAGATION IN NONLINEAR OPTICS: SUPPLEMENTAL DOCUMENT

To properly understand the rank 4 susceptibility tensor composed of 81 terms, the third order polarization vector equations and to understand the following sections, index notation is introduced.

### Section 1: The Polarization Vector and Rank 4 Susceptibility Tensor in Cartesian and Cylindrical Coordinates Using Index Notation

The nonlinear polarization vector equation is re-written using index notion $(i, j, k, l = x, y, z)$ [1], while explicitly showing the summations for added clarity,

$$P_i^{\mathrm{NL}}(\mathbf{r},t) = \varepsilon_0 \sum_j \sum_k \sum_l E_j(\mathbf{r},t) \int_{-\infty}^{\infty} \left[ \chi_{ijkl}^{(3)}(t-t') E_k(\mathbf{r},t') E_l(\mathbf{r},t') \right] dt', \tag{S1}$$

where in Cartesian coordinates $i = x,y,z$. The nonlinear susceptibility, $\chi_{ijkl}^{(3)}$, is the summation of the electronic and the nuclear contributions, $\chi_{ijkl}^{(3)}(t) = \chi_{\mathrm{el},ijkl}^{(3)}(t) + \chi_{\mathrm{nu},ijkl}^{(3)}(t)$, composed of a summation of rank 4 isotropic tensors $\delta_{ij}\delta_{kl}$ and $(\delta_{ik}\delta_{jl} + \delta_{il}\delta_{jk})$,

$$\begin{aligned} \chi_{ijkl}^{(3)}(t) &= \chi_0^{(3)} g_{ijkl}(t) \\ &= \chi_0^{(3)} \left[ \frac{f_{\mathrm{el}}}{3} \delta(t) \left( \delta_{ij}\delta_{kl} + \delta_{ik}\delta_{jl} + \delta_{il}\delta_{jk} \right) \right. \\ &\quad \left. + f_{\mathrm{nu,a}} g_{\mathrm{nu,a}}(t) \delta_{ij}\delta_{kl} + \frac{1}{2} f_{\mathrm{nu,b}} g_{\mathrm{nu,b}}(t) \left( \delta_{ik}\delta_{jl} + \delta_{il}\delta_{jk} \right) \right], \end{aligned} \tag{S2}$$

where $g_{xxxx}(t) = f_{\mathrm{el}}\delta(t) + f_{\mathrm{nu,a}} g_{\mathrm{nu,a}}(t) + f_{\mathrm{nu,b}} g_{\mathrm{nu,b}}(t)$. The response functions are normalized such that $\int_0^\infty g_{\mathrm{nu,a}}(t)dt = 1$ and $\int_0^\infty g_{\mathrm{nu,b}}(t)dt = 1$ and $\int_0^\infty g_{xxxx}(t)dt = 1$ matching [1-3]. This is discussed further in Supplement 1 of [4]. The rank 4 susceptibility tensor $\boldsymbol{\chi}^{(3)}(t)$ is related to the index notation $\chi_{ijkl}^{(3)}(t)$ using the summation of the unit vector dyadic products, dyads, $\hat{\boldsymbol{a}}_i\hat{\boldsymbol{a}}_j$ and $\hat{\boldsymbol{a}}_k\hat{\boldsymbol{a}}_l$ with $\chi_{ijkl}^{(3)}(t)$,

$$\boldsymbol{\chi}^{(3)}(t) = \sum_i \sum_j \sum_k \sum_l \hat{\boldsymbol{a}}_i \hat{\boldsymbol{a}}_j \, \chi_{ijkl}^{(3)}(t) \, \hat{\boldsymbol{a}}_k \hat{\boldsymbol{a}}_l \ . \tag{S3}$$

For this paper we are using $i = r, \phi, z$, requiring the 81 component rank 4 tensor $\chi_{ijkl}^{(3)}$, measured in Cartesian coordinates, to be transformed into cylindrical coordinates. Thankfully, by definition of a Cartesian isotropic tensor, the Cartesian isotropic tensor can be rotated at any point from $\hat{\boldsymbol{a}}_x$, $\hat{\boldsymbol{a}}_y$, $\hat{\boldsymbol{a}}_z$ to $\hat{\boldsymbol{a}}'_x$, $\hat{\boldsymbol{a}}'_y$, $\hat{\boldsymbol{a}}'_z$ without the isotropic tensor components, $\chi_{ijkl}^{(3)}$ changing [5-7]. For an isotropic rank 4 tensor, the transformation from Cartesian to Cylindrical coordinates, in the case of the unitless nonlinear susceptibility tensor, is equivalent to rotating the cartesian unit vectors $\hat{\boldsymbol{a}}_x$ and $\hat{\boldsymbol{a}}_y$ at the point $(x_p, y_p, z_p)$ to be parallel with the cylindrical unit vectors $\hat{\boldsymbol{a}}_r$ and $\hat{\boldsymbol{a}}_\phi$ at the point $(x_p, y_p, z_p)$. In this case, the transformed coordinates $\hat{\boldsymbol{a}}'_x$, $\hat{\boldsymbol{a}}'_y$, $\hat{\boldsymbol{a}}'_z$, become $\hat{\boldsymbol{a}}'_x = \hat{\boldsymbol{a}}_r$ , $\hat{\boldsymbol{a}}'_y = \hat{\boldsymbol{a}}_\phi$, $\hat{\boldsymbol{a}}'_z = \hat{\boldsymbol{a}}_z$. For visual purposes, rotation at point $(x_p, y_p, z_p)$, and transformation information is illustrated in Fig. 8. The result of this discussion is that the Cartesian susceptibility tensor components $\chi_{ijkl}^{(3)}$ for an isotropic media, can be used for the Cylindrical $\boldsymbol{\chi}^{(3)}$ without the rank 4 tensor components changing, making performing the transformation calculations unnecessary.

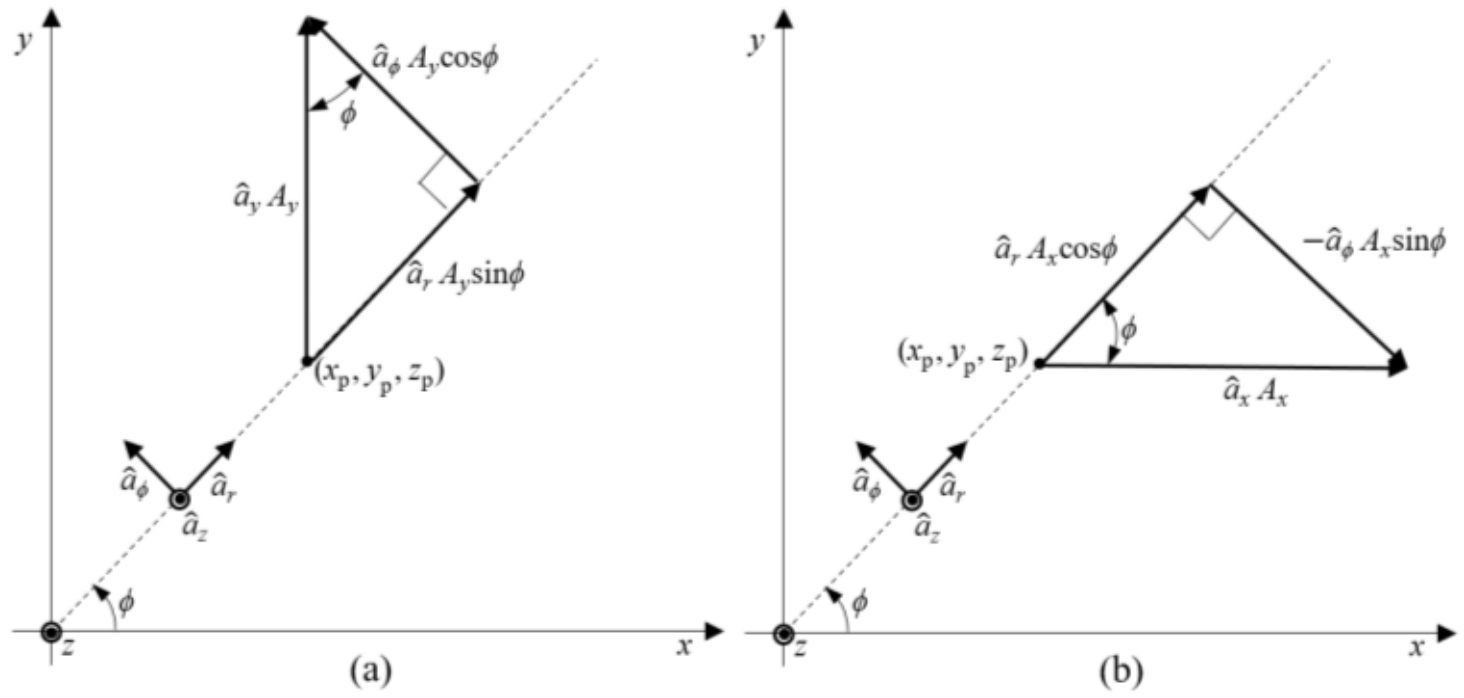

a. Fig. 8. The cartesian to cylindrical transformation/rotation is illustrated, where $\hat{\boldsymbol{a}}_x = \hat{\boldsymbol{a}}_r\cos\phi - \hat{\boldsymbol{a}}_\phi\sin\phi$, $\hat{\boldsymbol{a}}_y = \hat{\boldsymbol{a}}_r\sin\phi + \hat{\boldsymbol{a}}_\phi\cos\phi$ and $\hat{\boldsymbol{a}}_z = \hat{\boldsymbol{a}}_z$. This figure was recreated and modified from Figure VII.2 in [8], to better visualize the relationship between Cylindrical and Cartesian coordinates.

However, the transformation is shown below in detail for added clarity. Note that there are three different tensor terms, $\delta_{ij}\delta_{kl}$, $\delta_{ik}\delta_{jl}$ and $\delta_{il}\delta_{jk}$ which need to be transformed from Cartesian to Cylindrical, which are multiplied by the independent $\chi^{(3)}_{xxyy}(t)$, $\chi^{(3)}_{xyxy}(t)$ and $\chi^{(3)}_{xyyx}(t)$ respectively. In this case $\chi^{(3)}_{xyxy}(t) = \chi^{(3)}_{xyyx}(t)$ and $\chi^{(3)}_{xxxx}(t) = \chi^{(3)}_{yyyy}(t) = \chi^{(3)}_{zzzz}(t) = \chi^{(3)}_{xxyy}(t) + \chi^{(3)}_{xyxy}(t) + \chi^{(3)}_{xyyx}(t) = \chi^{(3)}_{xxyy}(t) + 2\chi^{(3)}_{xyxy}(t)$ [1].

The summation terms resulting from $\delta_{ij}\delta_{kl}$ are:

$$\begin{aligned}\hat{\boldsymbol{a}}_x\hat{\boldsymbol{a}}_x\,\chi^{(3)}_{xxyy}(t)\hat{\boldsymbol{a}}_x\hat{\boldsymbol{a}}_x + \hat{\boldsymbol{a}}_y\hat{\boldsymbol{a}}_y\,\chi^{(3)}_{xxyy}(t)\,\hat{\boldsymbol{a}}_y\hat{\boldsymbol{a}}_y + \hat{\boldsymbol{a}}_z\hat{\boldsymbol{a}}_z\,\chi^{(3)}_{xxyy}(t)\hat{\boldsymbol{a}}_z\hat{\boldsymbol{a}}_z + \hat{\boldsymbol{a}}_x\hat{\boldsymbol{a}}_x\,\chi^{(3)}_{xxyy}(t)\,\hat{\boldsymbol{a}}_y\hat{\boldsymbol{a}}_y \\ + \hat{\boldsymbol{a}}_x\hat{\boldsymbol{a}}_x\,\chi^{(3)}_{xxyy}(t)\,\hat{\boldsymbol{a}}_z\hat{\boldsymbol{a}}_z + \hat{\boldsymbol{a}}_y\hat{\boldsymbol{a}}_y\,\chi^{(3)}_{xxyy}(t)\,\hat{\boldsymbol{a}}_x\hat{\boldsymbol{a}}_x + \hat{\boldsymbol{a}}_z\hat{\boldsymbol{a}}_z\,\chi^{(3)}_{xxyy}(t)\,\hat{\boldsymbol{a}}_x\hat{\boldsymbol{a}}_x \\ + \hat{\boldsymbol{a}}_y\hat{\boldsymbol{a}}_y\,\chi^{(3)}_{xxyy}(t)\,\hat{\boldsymbol{a}}_z\hat{\boldsymbol{a}}_z + \hat{\boldsymbol{a}}_z\hat{\boldsymbol{a}}_z\,\chi^{(3)}_{xxyy}(t)\,\hat{\boldsymbol{a}}_y\hat{\boldsymbol{a}}_y\,,\end{aligned} \tag{S4}$$

$$\begin{aligned}\chi^{(3)}_{xxyy}(t)\big[\hat{\boldsymbol{a}}_x\hat{\boldsymbol{a}}_x\hat{\boldsymbol{a}}_x\hat{\boldsymbol{a}}_x + \hat{\boldsymbol{a}}_y\hat{\boldsymbol{a}}_y\hat{\boldsymbol{a}}_y\hat{\boldsymbol{a}}_y + \hat{\boldsymbol{a}}_z\hat{\boldsymbol{a}}_z\hat{\boldsymbol{a}}_z\hat{\boldsymbol{a}}_z + \hat{\boldsymbol{a}}_x\hat{\boldsymbol{a}}_x\hat{\boldsymbol{a}}_y\hat{\boldsymbol{a}}_y + \hat{\boldsymbol{a}}_x\hat{\boldsymbol{a}}_x\hat{\boldsymbol{a}}_z\hat{\boldsymbol{a}}_z + \hat{\boldsymbol{a}}_y\hat{\boldsymbol{a}}_y\hat{\boldsymbol{a}}_x\hat{\boldsymbol{a}}_x \\ + \hat{\boldsymbol{a}}_z\hat{\boldsymbol{a}}_z\hat{\boldsymbol{a}}_x\hat{\boldsymbol{a}}_x + \hat{\boldsymbol{a}}_y\hat{\boldsymbol{a}}_y\hat{\boldsymbol{a}}_z\hat{\boldsymbol{a}}_z + \hat{\boldsymbol{a}}_z\hat{\boldsymbol{a}}_z\hat{\boldsymbol{a}}_y\hat{\boldsymbol{a}}_y\big]\,.\end{aligned} \tag{S5}$$

The summation terms resulting from $\delta_{ik}\delta_{jl}$ are:

$$\begin{aligned}\hat{\boldsymbol{a}}_x\hat{\boldsymbol{a}}_x\,\chi^{(3)}_{xyxy}(t)\hat{\boldsymbol{a}}_x\hat{\boldsymbol{a}}_x + \hat{\boldsymbol{a}}_y\hat{\boldsymbol{a}}_y\,\chi^{(3)}_{xyxy}(t)\,\hat{\boldsymbol{a}}_y\hat{\boldsymbol{a}}_y + \hat{\boldsymbol{a}}_z\hat{\boldsymbol{a}}_z\,\chi^{(3)}_{xyxy}(t)\,\hat{\boldsymbol{a}}_z\hat{\boldsymbol{a}}_z + \hat{\boldsymbol{a}}_x\hat{\boldsymbol{a}}_y\,\chi^{(3)}_{xyxy}(t)\,\hat{\boldsymbol{a}}_x\hat{\boldsymbol{a}}_y \\ + \hat{\boldsymbol{a}}_x\hat{\boldsymbol{a}}_z\,\chi^{(3)}_{xyxy}(t)\,\hat{\boldsymbol{a}}_x\hat{\boldsymbol{a}}_z + \hat{\boldsymbol{a}}_y\hat{\boldsymbol{a}}_x\,\chi^{(3)}_{xyxy}(t)\,\hat{\boldsymbol{a}}_y\hat{\boldsymbol{a}}_x + \hat{\boldsymbol{a}}_z\hat{\boldsymbol{a}}_x\,\chi^{(3)}_{xyxy}(t)\,\hat{\boldsymbol{a}}_z\hat{\boldsymbol{a}}_x \\ + \hat{\boldsymbol{a}}_y\hat{\boldsymbol{a}}_z\,\chi^{(3)}_{xyxy}(t)\,\hat{\boldsymbol{a}}_y\hat{\boldsymbol{a}}_z + \hat{\boldsymbol{a}}_z\hat{\boldsymbol{a}}_y\,\chi^{(3)}_{xyxy}(t)\,\hat{\boldsymbol{a}}_z\hat{\boldsymbol{a}}_y\,,\end{aligned} \tag{S6}$$

$$\begin{aligned}\chi^{(3)}_{xyxy}(t)\big[\hat{\boldsymbol{a}}_x\hat{\boldsymbol{a}}_x\hat{\boldsymbol{a}}_x\hat{\boldsymbol{a}}_x + \hat{\boldsymbol{a}}_y\hat{\boldsymbol{a}}_y\hat{\boldsymbol{a}}_y\hat{\boldsymbol{a}}_y + \hat{\boldsymbol{a}}_z\hat{\boldsymbol{a}}_z\hat{\boldsymbol{a}}_z\hat{\boldsymbol{a}}_z + \hat{\boldsymbol{a}}_x\hat{\boldsymbol{a}}_y\hat{\boldsymbol{a}}_x\hat{\boldsymbol{a}}_y + \hat{\boldsymbol{a}}_x\hat{\boldsymbol{a}}_z\hat{\boldsymbol{a}}_x\hat{\boldsymbol{a}}_z + \hat{\boldsymbol{a}}_y\hat{\boldsymbol{a}}_x\hat{\boldsymbol{a}}_y\hat{\boldsymbol{a}}_x \\ + \hat{\boldsymbol{a}}_z\hat{\boldsymbol{a}}_x\hat{\boldsymbol{a}}_z\hat{\boldsymbol{a}}_x + \hat{\boldsymbol{a}}_y\hat{\boldsymbol{a}}_z\hat{\boldsymbol{a}}_y\hat{\boldsymbol{a}}_z + \hat{\boldsymbol{a}}_z\hat{\boldsymbol{a}}_y\hat{\boldsymbol{a}}_z\hat{\boldsymbol{a}}_y\big]\,.\end{aligned} \tag{S7}$$

The summation terms resulting from $\delta_{il}\delta_{jk}$ are:

$$\begin{aligned}\hat{\boldsymbol{a}}_x\hat{\boldsymbol{a}}_x\,\chi^{(3)}_{xyyx}(t)\,\hat{\boldsymbol{a}}_x\hat{\boldsymbol{a}}_x + \hat{\boldsymbol{a}}_y\hat{\boldsymbol{a}}_y\,\chi^{(3)}_{xyyx}(t)\,\hat{\boldsymbol{a}}_y\hat{\boldsymbol{a}}_y + \hat{\boldsymbol{a}}_z\hat{\boldsymbol{a}}_z\,\chi^{(3)}_{xyyx}(t)\,\hat{\boldsymbol{a}}_z\hat{\boldsymbol{a}}_z + \hat{\boldsymbol{a}}_x\hat{\boldsymbol{a}}_y\,\chi^{(3)}_{xyyx}(t)\,\hat{\boldsymbol{a}}_y\hat{\boldsymbol{a}}_x \\ + \hat{\boldsymbol{a}}_x\hat{\boldsymbol{a}}_z\,\chi^{(3)}_{xyyx}(t)\,\hat{\boldsymbol{a}}_z\hat{\boldsymbol{a}}_x + \hat{\boldsymbol{a}}_y\hat{\boldsymbol{a}}_x\,\chi^{(3)}_{xyyx}(t)\,\hat{\boldsymbol{a}}_x\hat{\boldsymbol{a}}_y + \hat{\boldsymbol{a}}_z\hat{\boldsymbol{a}}_x\,\chi^{(3)}_{xyyx}(t)\,\hat{\boldsymbol{a}}_x\hat{\boldsymbol{a}}_z \\ + \hat{\boldsymbol{a}}_y\hat{\boldsymbol{a}}_z\,\chi^{(3)}_{xyyx}(t)\,\hat{\boldsymbol{a}}_z\hat{\boldsymbol{a}}_y + \hat{\boldsymbol{a}}_z\hat{\boldsymbol{a}}_y\,\chi^{(3)}_{xyyx}(t)\,\hat{\boldsymbol{a}}_y\hat{\boldsymbol{a}}_z\,,\end{aligned} \tag{S8}$$

$$\begin{aligned}\chi^{(3)}_{xyyx}(t)\big[\hat{\boldsymbol{a}}_x\hat{\boldsymbol{a}}_x\hat{\boldsymbol{a}}_x\hat{\boldsymbol{a}}_x + \hat{\boldsymbol{a}}_y\hat{\boldsymbol{a}}_y\hat{\boldsymbol{a}}_y\hat{\boldsymbol{a}}_y + \hat{\boldsymbol{a}}_z\hat{\boldsymbol{a}}_z\hat{\boldsymbol{a}}_z\hat{\boldsymbol{a}}_z + \hat{\boldsymbol{a}}_x\hat{\boldsymbol{a}}_y\hat{\boldsymbol{a}}_y\hat{\boldsymbol{a}}_x + \hat{\boldsymbol{a}}_x\hat{\boldsymbol{a}}_z\hat{\boldsymbol{a}}_z\hat{\boldsymbol{a}}_x + \hat{\boldsymbol{a}}_y\hat{\boldsymbol{a}}_x\hat{\boldsymbol{a}}_x\hat{\boldsymbol{a}}_y \\ + \hat{\boldsymbol{a}}_z\hat{\boldsymbol{a}}_x\hat{\boldsymbol{a}}_x\hat{\boldsymbol{a}}_z + \hat{\boldsymbol{a}}_y\hat{\boldsymbol{a}}_z\hat{\boldsymbol{a}}_z\hat{\boldsymbol{a}}_y + \hat{\boldsymbol{a}}_z\hat{\boldsymbol{a}}_y\hat{\boldsymbol{a}}_y\hat{\boldsymbol{a}}_z\big]\,.\end{aligned} \tag{S9}$$

To transform these terms from cartesian coordinates to cylindrical coordinates, $\hat{\boldsymbol{a}}_x = \hat{\boldsymbol{a}}_r\cos\phi - \hat{\boldsymbol{a}}_\phi\sin\phi$, $\hat{\boldsymbol{a}}_y = \hat{\boldsymbol{a}}_r\sin\phi + \hat{\boldsymbol{a}}_\phi\cos\phi$ and $\hat{\boldsymbol{a}}_z = \hat{\boldsymbol{a}}_z$ are plugged into the dyads as shown below.

$$\hat{\boldsymbol{a}}_x\hat{\boldsymbol{a}}_x = (\hat{\boldsymbol{a}}_r\cos\phi - \hat{\boldsymbol{a}}_\phi\sin\phi)(\hat{\boldsymbol{a}}_r\cos\phi - \hat{\boldsymbol{a}}_\phi\sin\phi) = \hat{\boldsymbol{a}}_r\hat{\boldsymbol{a}}_r(\cos^2\phi) + \hat{\boldsymbol{a}}_\phi\hat{\boldsymbol{a}}_\phi(\sin^2\phi) - \hat{\boldsymbol{a}}_r\hat{\boldsymbol{a}}_\phi(\cos\phi\sin\phi) - \hat{\boldsymbol{a}}_\phi\hat{\boldsymbol{a}}_r(\cos\phi\sin\phi)\,, \tag{S10}$$

$$\hat{\boldsymbol{a}}_y\hat{\boldsymbol{a}}_y = (\hat{\boldsymbol{a}}_r\sin\phi + \hat{\boldsymbol{a}}_\phi\cos\phi)(\hat{\boldsymbol{a}}_r\sin\phi + \hat{\boldsymbol{a}}_\phi\cos\phi) = \hat{\boldsymbol{a}}_r\hat{\boldsymbol{a}}_r(\sin^2\phi) + \hat{\boldsymbol{a}}_\phi\hat{\boldsymbol{a}}_\phi(\cos^2\phi) + \hat{\boldsymbol{a}}_r\hat{\boldsymbol{a}}_\phi(\cos\phi\sin\phi) + \hat{\boldsymbol{a}}_\phi\hat{\boldsymbol{a}}_r(\cos\phi\sin\phi)\,, \tag{S11}$$

$$\hat{\boldsymbol{a}}_x\hat{\boldsymbol{a}}_y = (\hat{\boldsymbol{a}}_r\cos\phi - \hat{\boldsymbol{a}}_\phi\sin\phi)(\hat{\boldsymbol{a}}_r\sin\phi + \hat{\boldsymbol{a}}_\phi\cos\phi) = \hat{\boldsymbol{a}}_r\hat{\boldsymbol{a}}_r(\cos\phi\sin\phi) - \hat{\boldsymbol{a}}_\phi\hat{\boldsymbol{a}}_\phi(\cos\phi\sin\phi) + \hat{\boldsymbol{a}}_r\hat{\boldsymbol{a}}_\phi(\cos^2\phi) - \hat{\boldsymbol{a}}_\phi\hat{\boldsymbol{a}}_r(\sin^2\phi)\,, \tag{S12}$$

$$\hat{\boldsymbol{a}}_y\hat{\boldsymbol{a}}_x = (\hat{\boldsymbol{a}}_r\sin\phi + \hat{\boldsymbol{a}}_\phi\cos\phi)(\hat{\boldsymbol{a}}_r\cos\phi - \hat{\boldsymbol{a}}_\phi\sin\phi) = \hat{\boldsymbol{a}}_r\hat{\boldsymbol{a}}_r(\cos\phi\sin\phi) - \hat{\boldsymbol{a}}_\phi\hat{\boldsymbol{a}}_\phi(\cos\phi\sin\phi) - \hat{\boldsymbol{a}}_r\hat{\boldsymbol{a}}_\phi(\sin^2\phi) + \hat{\boldsymbol{a}}_\phi\hat{\boldsymbol{a}}_r(\cos^2\phi)\,. \tag{S13}$$

Next, the dyads are plugged into the $\delta_{ij}\delta_{kl}$ tetrads:

$$\hat{\boldsymbol{a}}_x\hat{\boldsymbol{a}}_x\hat{\boldsymbol{a}}_x\hat{\boldsymbol{a}}_x = \left[\hat{\boldsymbol{a}}_r\hat{\boldsymbol{a}}_r(\cos^2\phi) + \hat{\boldsymbol{a}}_\phi\hat{\boldsymbol{a}}_\phi(\sin^2\phi) - \hat{\boldsymbol{a}}_r\hat{\boldsymbol{a}}_\phi(\cos\phi\sin\phi) - \hat{\boldsymbol{a}}_\phi\hat{\boldsymbol{a}}_r(\cos\phi\sin\phi)\right]^2, \tag{S14}$$

$$\hat{\boldsymbol{a}}_y\hat{\boldsymbol{a}}_y\hat{\boldsymbol{a}}_y\hat{\boldsymbol{a}}_y = \left[\hat{\boldsymbol{a}}_r\hat{\boldsymbol{a}}_r(\sin^2\phi) + \hat{\boldsymbol{a}}_\phi\hat{\boldsymbol{a}}_\phi(\cos^2\phi) + \hat{\boldsymbol{a}}_r\hat{\boldsymbol{a}}_\phi(\cos\phi\sin\phi) + \hat{\boldsymbol{a}}_\phi\hat{\boldsymbol{a}}_r(\cos\phi\sin\phi)\right]^2, \tag{S15}$$

$$\hat{\boldsymbol{a}}_x\hat{\boldsymbol{a}}_x\hat{\boldsymbol{a}}_y\hat{\boldsymbol{a}}_y = \left[\hat{\boldsymbol{a}}_r\hat{\boldsymbol{a}}_r(\cos^2\phi) + \hat{\boldsymbol{a}}_\phi\hat{\boldsymbol{a}}_\phi(\sin^2\phi) - \hat{\boldsymbol{a}}_r\hat{\boldsymbol{a}}_\phi(\cos\phi\sin\phi) - \hat{\boldsymbol{a}}_\phi\hat{\boldsymbol{a}}_r(\cos\phi\sin\phi)\right] \cdot \left[\hat{\boldsymbol{a}}_r\hat{\boldsymbol{a}}_r(\sin^2\phi) + \hat{\boldsymbol{a}}_\phi\hat{\boldsymbol{a}}_\phi(\cos^2\phi) + \hat{\boldsymbol{a}}_r\hat{\boldsymbol{a}}_\phi(\cos\phi\sin\phi) + \hat{\boldsymbol{a}}_\phi\hat{\boldsymbol{a}}_r(\cos\phi\sin\phi)\right], \tag{S16}$$

$$\hat{\boldsymbol{a}}_y\hat{\boldsymbol{a}}_y\hat{\boldsymbol{a}}_x\hat{\boldsymbol{a}}_x = \left[\hat{\boldsymbol{a}}_r\hat{\boldsymbol{a}}_r(\sin^2\phi) + \hat{\boldsymbol{a}}_\phi\hat{\boldsymbol{a}}_\phi(\cos^2\phi) + \hat{\boldsymbol{a}}_r\hat{\boldsymbol{a}}_\phi(\cos\phi\sin\phi) + \hat{\boldsymbol{a}}_\phi\hat{\boldsymbol{a}}_r(\cos\phi\sin\phi)\right] \cdot \left[\hat{\boldsymbol{a}}_r\hat{\boldsymbol{a}}_r(\cos^2\phi) + \hat{\boldsymbol{a}}_\phi\hat{\boldsymbol{a}}_\phi(\sin^2\phi) - \hat{\boldsymbol{a}}_r\hat{\boldsymbol{a}}_\phi(\cos\phi\sin\phi) - \hat{\boldsymbol{a}}_\phi\hat{\boldsymbol{a}}_r(\cos\phi\sin\phi)\right], \tag{S17}$$

$$\hat{\boldsymbol{a}}_x\hat{\boldsymbol{a}}_x\hat{\boldsymbol{a}}_z\hat{\boldsymbol{a}}_z = \left[\hat{\boldsymbol{a}}_r\hat{\boldsymbol{a}}_r(\cos^2\phi) + \hat{\boldsymbol{a}}_\phi\hat{\boldsymbol{a}}_\phi(\sin^2\phi) - \hat{\boldsymbol{a}}_r\hat{\boldsymbol{a}}_\phi(\cos\phi\sin\phi) - \hat{\boldsymbol{a}}_\phi\hat{\boldsymbol{a}}_r(\cos\phi\sin\phi)\right]\hat{\boldsymbol{a}}_z\hat{\boldsymbol{a}}_z\,, \tag{S18}$$

$$\hat{\boldsymbol{a}}_y\hat{\boldsymbol{a}}_y\hat{\boldsymbol{a}}_z\hat{\boldsymbol{a}}_z = [\hat{\boldsymbol{a}}_r\hat{\boldsymbol{a}}_r(\sin^2\phi) + \hat{\boldsymbol{a}}_\phi\hat{\boldsymbol{a}}_\phi(\cos^2\phi) + \hat{\boldsymbol{a}}_r\hat{\boldsymbol{a}}_\phi(\cos\phi\sin\phi) + \hat{\boldsymbol{a}}_\phi\hat{\boldsymbol{a}}_r(\cos\phi\sin\phi)]\hat{\boldsymbol{a}}_z\hat{\boldsymbol{a}}_z\,, \tag{S19}$$

$$\hat{\boldsymbol{a}}_z\hat{\boldsymbol{a}}_z\hat{\boldsymbol{a}}_x\hat{\boldsymbol{a}}_x = \hat{\boldsymbol{a}}_z\hat{\boldsymbol{a}}_z\left[\hat{\boldsymbol{a}}_r\hat{\boldsymbol{a}}_r(\cos^2\phi) + \hat{\boldsymbol{a}}_\phi\hat{\boldsymbol{a}}_\phi(\sin^2\phi) - \hat{\boldsymbol{a}}_r\hat{\boldsymbol{a}}_\phi(\cos\phi\sin\phi) - \hat{\boldsymbol{a}}_\phi\hat{\boldsymbol{a}}_r(\cos\phi\sin\phi)\right], \tag{S20}$$

$$\hat{\boldsymbol{a}}_z\hat{\boldsymbol{a}}_z\hat{\boldsymbol{a}}_y\hat{\boldsymbol{a}}_y = \hat{\boldsymbol{a}}_z\hat{\boldsymbol{a}}_z\left[\hat{\boldsymbol{a}}_r\hat{\boldsymbol{a}}_r(\sin^2\phi) + \hat{\boldsymbol{a}}_\phi\hat{\boldsymbol{a}}_\phi(\cos^2\phi) + \hat{\boldsymbol{a}}_r\hat{\boldsymbol{a}}_\phi(\cos\phi\sin\phi) + \hat{\boldsymbol{a}}_\phi\hat{\boldsymbol{a}}_r(\cos\phi\sin\phi)\right]. \tag{S21}$$

Adding the above terms, where the negative terms cancel terms from other tetrads, and using the identities $\sin^2\phi + \cos^2\phi = 1$ and $\sin^4\phi + \cos^4\phi + 2\sin^2\phi\cos^2\phi = 1$.

$$\chi^{(3)}_{xxyy}(t)\big[\hat{a}_x\hat{a}_x\hat{a}_x\hat{a}_x + \hat{a}_y\hat{a}_y\hat{a}_y\hat{a}_y + \hat{a}_z\hat{a}_z\hat{a}_z\hat{a}_z + \hat{a}_x\hat{a}_x\hat{a}_y\hat{a}_y + \hat{a}_x\hat{a}_x\hat{a}_z\hat{a}_z + \hat{a}_y\hat{a}_y\hat{a}_x\hat{a}_x + \hat{a}_z\hat{a}_z\hat{a}_x\hat{a}_x + \hat{a}_y\hat{a}_y\hat{a}_z\hat{a}_z + \hat{a}_z\hat{a}_z\hat{a}_y\hat{a}_y\big] = \chi^{(3)}_{xxyy}(t)\big[\hat{a}_r\hat{a}_r\hat{a}_r\hat{a}_r + \hat{a}_\phi\hat{a}_\phi\hat{a}_\phi\hat{a}_\phi + \hat{a}_z\hat{a}_z\hat{a}_z\hat{a}_z + \hat{a}_r\hat{a}_r\hat{a}_\phi\hat{a}_\phi + \hat{a}_r\hat{a}_r\hat{a}_z\hat{a}_z + \hat{a}_\phi\hat{a}_\phi\hat{a}_r\hat{a}_r + \hat{a}_z\hat{a}_z\hat{a}_r\hat{a}_r + \hat{a}_\phi\hat{a}_\phi\hat{a}_z\hat{a}_z + \hat{a}_z\hat{a}_z\hat{a}_\phi\hat{a}_\phi\big]. \tag{S22}$$

This confirms that the cylindrical coordinates tensor has the same component value of $\chi^{(3)}_{xxyy}(t)$ for the $\delta_{ij}\delta_{kl}$ components, meaning $\chi^{(3)}_{rr\phi\phi}(t) = \chi^{(3)}_{xxyy}(t)$. Next, the dyads are plugged into the $\delta_{ik}\delta_{jl}$ tetrads:

$$\hat{a}_x\hat{a}_y\hat{a}_x\hat{a}_y = \big[\hat{a}_r\hat{a}_r(\cos\phi\sin\phi) - \hat{a}_\phi\hat{a}_\phi(\cos\phi\sin\phi) + \hat{a}_r\hat{a}_\phi(\cos^2\phi) - \hat{a}_\phi\hat{a}_r(\sin^2\phi)\big]^2, \tag{S23}$$

$$\hat{a}_y\hat{a}_x\hat{a}_y\hat{a}_x = \big[\hat{a}_r\hat{a}_r(\cos\phi\sin\phi) - \hat{a}_\phi\hat{a}_\phi(\cos\phi\sin\phi) - \hat{a}_r\hat{a}_\phi(\sin^2\phi) + \hat{a}_\phi\hat{a}_r(\cos^2\phi)\big]^2, \tag{S24}$$

$$\hat{a}_x\hat{a}_z\hat{a}_x\hat{a}_z = \big[\hat{a}_r\hat{a}_z\cos\phi - \hat{a}_\phi\hat{a}_z\sin\phi\big]\big[\hat{a}_r\hat{a}_z\cos\phi - \hat{a}_\phi\hat{a}_z\sin\phi\big], \tag{S25}$$

$$\hat{a}_y\hat{a}_z\hat{a}_y\hat{a}_z = \big[\hat{a}_r\hat{a}_z\sin\phi + \hat{a}_\phi\hat{a}_z\cos\phi\big]\big[\hat{a}_r\hat{a}_z\sin\phi + \hat{a}_\phi\hat{a}_z\cos\phi\big], \tag{S26}$$

$$\hat{a}_z\hat{a}_x\hat{a}_z\hat{a}_x = \big[\hat{a}_z\hat{a}_r\cos\phi - \hat{a}_z\hat{a}_\phi\sin\phi\big]\big[\hat{a}_z\hat{a}_r\cos\phi - \hat{a}_z\hat{a}_\phi\sin\phi\big], \tag{S27}$$

$$\hat{a}_z\hat{a}_y\hat{a}_z\hat{a}_y = \big[\hat{a}_z\hat{a}_r\sin\phi + \hat{a}_z\hat{a}_\phi\cos\phi\big]\big[\hat{a}_z\hat{a}_r\sin\phi + \hat{a}_z\hat{a}_\phi\cos\phi\big]. \tag{S28}$$

Adding the above terms, where the negative terms cancel terms from other tetrads, and using the identities $\sin^2\phi + \cos^2\phi = 1$ and $\sin^4\phi + \cos^4\phi + 2\sin^2\phi\cos^2\phi = 1$.

$$\chi^{(3)}_{xyxy}(t)\big[\hat{a}_x\hat{a}_x\hat{a}_x\hat{a}_x + \hat{a}_y\hat{a}_y\hat{a}_y\hat{a}_y + \hat{a}_z\hat{a}_z\hat{a}_z\hat{a}_z + \hat{a}_x\hat{a}_y\hat{a}_x\hat{a}_y + \hat{a}_x\hat{a}_z\hat{a}_x\hat{a}_z + \hat{a}_y\hat{a}_x\hat{a}_y\hat{a}_x + \hat{a}_z\hat{a}_x\hat{a}_z\hat{a}_x + \hat{a}_y\hat{a}_z\hat{a}_y\hat{a}_z + \hat{a}_z\hat{a}_y\hat{a}_z\hat{a}_y\big] = \chi^{(3)}_{xyxy}(t)\big[\hat{a}_r\hat{a}_r\hat{a}_r\hat{a}_r + \hat{a}_\phi\hat{a}_\phi\hat{a}_\phi\hat{a}_\phi + \hat{a}_z\hat{a}_z\hat{a}_z\hat{a}_z + \hat{a}_r\hat{a}_y\hat{a}_r\hat{a}_y + \hat{a}_r\hat{a}_z\hat{a}_r\hat{a}_z + \hat{a}_\phi\hat{a}_r\hat{a}_\phi\hat{a}_r + \hat{a}_z\hat{a}_r\hat{a}_z\hat{a}_r + \hat{a}_\phi\hat{a}_z\hat{a}_\phi\hat{a}_z + \hat{a}_z\hat{a}_\phi\hat{a}_z\hat{a}_\phi\big]. \tag{S29}$$

This confirms that the cylindrical coordinates tensor has the same component value of $\chi^{(3)}_{xyxy}(t)$ for the $\delta_{ik}\delta_{jl}$ components, meaning $\chi^{(3)}_{r\phi r\phi}(t) = \chi^{(3)}_{xyxy}(t)$. Next, the dyads are plugged into the $\delta_{il}\delta_{jk}$ tetrads:

$$\hat{a}_x\hat{a}_y\hat{a}_y\hat{a}_x = \big[\hat{a}_r\hat{a}_r(\cos\phi\sin\phi) - \hat{a}_\phi\hat{a}_\phi(\cos\phi\sin\phi) + \hat{a}_r\hat{a}_\phi(\cos^2\phi) - \hat{a}_\phi\hat{a}_r(\sin^2\phi)\big] \cdot \big[\hat{a}_r\hat{a}_r(\cos\phi\sin\phi) - \hat{a}_\phi\hat{a}_\phi(\cos\phi\sin\phi) - \hat{a}_r\hat{a}_\phi(\sin^2\phi) + \hat{a}_\phi\hat{a}_r(\cos^2\phi)\big], \tag{S30}$$

$$\hat{a}_y\hat{a}_x\hat{a}_x\hat{a}_y = \big[\hat{a}_r\hat{a}_r(\cos\phi\sin\phi) - \hat{a}_\phi\hat{a}_\phi(\cos\phi\sin\phi) - \hat{a}_r\hat{a}_\phi(\sin^2\phi) + \hat{a}_\phi\hat{a}_r(\cos^2\phi)\big] \cdot \big[\hat{a}_r\hat{a}_r(\cos\phi\sin\phi) - \hat{a}_\phi\hat{a}_\phi(\cos\phi\sin\phi) + \hat{a}_r\hat{a}_\phi(\cos^2\phi) - \hat{a}_\phi\hat{a}_r(\sin^2\phi)\big], \tag{S31}$$

$$\hat{a}_x\hat{a}_z\hat{a}_z\hat{a}_x = \big[\hat{a}_r\hat{a}_z\cos\phi - \hat{a}_\phi\hat{a}_z\sin\phi\big]\big[\hat{a}_z\hat{a}_r\cos\phi - \hat{a}_z\hat{a}_\phi\sin\phi\big], \tag{S32}$$

$$\hat{a}_y\hat{a}_z\hat{a}_z\hat{a}_y = \big[\hat{a}_r\hat{a}_z\sin\phi + \hat{a}_\phi\hat{a}_z\cos\phi\big]\big[\hat{a}_z\hat{a}_r\sin\phi + \hat{a}_z\hat{a}_\phi\cos\phi\big], \tag{S33}$$

$$\hat{a}_z\hat{a}_x\hat{a}_x\hat{a}_z = [\hat{a}_z\hat{a}_r\cos\phi - \hat{a}_z\hat{a}_\phi\sin\phi][\hat{a}_r\hat{a}_z\cos\phi - \hat{a}_\phi\hat{a}_z\sin\phi]\,, \tag{S34}$$

$$\hat{a}_z\hat{a}_y\hat{a}_y\hat{a}_z = [\hat{a}_z\hat{a}_r\sin\phi + \hat{a}_z\hat{a}_\phi\cos\phi][\hat{a}_r\hat{a}_z\sin\phi + \hat{a}_\phi\hat{a}_z\cos\phi]\,. \tag{S35}$$

Adding the above terms, where the negative terms cancel terms from other tetrads, and using the identities $\sin^2\phi + \cos^2\phi = 1$ and $\sin^4\phi + \cos^4\phi + 2\sin^2\phi\cos^2\phi = 1$.

$$\begin{aligned}\chi^{(3)}_{xyyx}(t)[\hat{a}_x\hat{a}_x\hat{a}_x\hat{a}_x + \hat{a}_y\hat{a}_y\hat{a}_y\hat{a}_y + \hat{a}_z\hat{a}_z\hat{a}_z\hat{a}_z + \hat{a}_x\hat{a}_y\hat{a}_y\hat{a}_x + \hat{a}_x\hat{a}_z\hat{a}_z\hat{a}_x + \hat{a}_y\hat{a}_x\hat{a}_x\hat{a}_y \\ + \hat{a}_z\hat{a}_x\hat{a}_x\hat{a}_z + \hat{a}_y\hat{a}_z\hat{a}_z\hat{a}_y + \hat{a}_z\hat{a}_y\hat{a}_y\hat{a}_z] \\ = \chi^{(3)}_{xyyx}(t)[\hat{a}_r\hat{a}_r\hat{a}_r\hat{a}_r + \hat{a}_\phi\hat{a}_\phi\hat{a}_\phi\hat{a}_\phi + \hat{a}_z\hat{a}_z\hat{a}_z\hat{a}_z + \hat{a}_r\hat{a}_\phi\hat{a}_\phi\hat{a}_r \\ + \hat{a}_r\hat{a}_z\hat{a}_z\hat{a}_r + \hat{a}_\phi\hat{a}_r\hat{a}_r\hat{a}_\phi + \hat{a}_z\hat{a}_r\hat{a}_r\hat{a}_z + \hat{a}_\phi\hat{a}_z\hat{a}_z\hat{a}_\phi \\ + \hat{a}_z\hat{a}_\phi\hat{a}_\phi\hat{a}_z]\,.\end{aligned} \tag{S36}$$

This confirms that the Cylindrical coordinates tensor has the same component value of $\chi^{(3)}_{xyyx}(t)$ for the $\delta_{il}\delta_{jk}$ components, meaning $\chi^{(3)}_{r\phi\phi r}(t) = \chi^{(3)}_{xyyx}(t)$.

Adding the $\delta_{ij}\delta_{kl}$, $\delta_{ik}\delta_{jl}$ and $\delta_{il}\delta_{jk}$ tetrads together from Eqns (S22), (S29), (S36) using $i = r, \phi, z$ yields $\boldsymbol{\chi}^{(3)}(t)$ in Cylindrical coordinates, where the $\chi^{(3)}_{ijkl}$ for Cylindrical coordinates can be seen to be the same as for Cartesian coordinates as predicated by the definition of a Cartesian isotropic tensor, where the tensor components are the same upon rotation at a given point in space, meaning $\chi^{(3)}_{rr\phi\phi}(t) = \chi^{(3)}_{xxyy}(t)$, $\chi^{(3)}_{r\phi r\phi}(t) = \chi^{(3)}_{xyxy}(t)$, $\chi^{(3)}_{r\phi\phi r}(t) = \chi^{(3)}_{xyyx}(t)$ and $\chi^{(3)}_{rrrr}(t) = \chi^{(3)}_{\phi\phi\phi\phi}(t) = \chi^{(3)}_{zzzz}(t) = \chi^{(3)}_{rr\phi\phi}(t) + \chi^{(3)}_{r\phi r\phi}(t) + \chi^{(3)}_{r\phi\phi r}(t) = \chi^{(3)}_{rr\phi\phi}(t) + 2\chi^{(3)}_{r\phi r\phi}(t)$.

## Section 2: Fused Silica Rank 4 Susceptibility Tensor

The normalized nonlinear rank 4 material response tensor for silica written using index notation is,

$$g^{(3)}_{ijkl}(t) = \left[\frac{(1 - f_{R,\mathrm{Isotropic}})}{3}\delta(t)(\delta_{ij}\delta_{kl} + \delta_{ik}\delta_{jl} + \delta_{il}\delta_{jk}) + f_{R,\mathrm{Isotropic}}\, g_{\mathrm{nu,a}}(t)\delta_{ij}\delta_{kl}\right]. \tag{S37}$$

## Section 3: Carbon Disulfide Rank 4 Susceptibility Tensor

The nonlinear rank 4 tensor for $CS_2$ written in a slightly modified form from [3] is,

$$g^{(3)}_{ijkl}(t) = \left[\frac{f_{\mathrm{el}}}{3}\delta(t)(\delta_{ij}\delta_{kl} + \delta_{ik}\delta_{jl} + \delta_{il}\delta_{jk}) + f_c\, g_c(t)\delta_{ij}\delta_{kl} + \frac{1}{2}[\, f_d\, g_d(t) + f_v\, g_v(t)](\delta_{ik}\delta_{jl} + \delta_{il}\delta_{jk})\right]. \tag{S38}$$

where $g_c(t)$, $g_v(t)$ and $g_d(t)$ correspond to the (overdamped) “collision”, (underdamped) "libration" resulting from molecular re-orientation, and (overdamped) “diffusive molecular re-orientation” nuclear response functions for $CS_2$ [9, 10].

The electronic and nuclear parts of the susceptibility can be calculated as $\chi^{(3)}_{\mathrm{el}} = (4/3)c\,\varepsilon_0\varepsilon_r n_{2,\mathrm{el}}$ and $\chi^{(3)}_{\mathrm{nu}}(t) = 2\,c\,\varepsilon_0\varepsilon_r\, R(t)$ where $R(t) = \sum_m n_{2,m} g_m(t)$ as described in [9, 11]. To determine $\chi^{(3)}_0$, $f_{\mathrm{el}}$, $f_c$, $f_l$ and $f_d$, the special case where $i = j = k = l = x$ was considered leading to,

$$\chi^{(3)}_{xxxx}(t) = (4/3)c\,\varepsilon_0\varepsilon_r n_{2,el} g_{el}(t) + 2\,c\,\varepsilon_0\varepsilon_r\big[n_{2,c}g_c(t) + n_{2,l}g_l(t) + n_{2,d}g_d(t)\big]. \tag{S39}$$

The equation can be re-written pulling out a constant $\chi_0^{(3)}$ as $\chi^{(3)}_{xxxx}(t) = \chi_0^{(3)} g_{xxxx}(t)$,

$$\chi^{(3)}_{xxxx}(t) = \chi_0^{(3)}[\,f_{el}g_{el}(t) + f_c g_c(t) + f_l g_l(t) + f_d g_d(t)\,], \tag{S40}$$

where $f_{el} = (2/3)n_{2,\mathrm{el}}/n_{2,\mathrm{sum}}$, $f_c = n_{2,c}/n_{2,\mathrm{sum}}$, $f_l = n_{2,l}/n_{2,\mathrm{sum}}$, $f_d = n_{2,d}/n_{2,\mathrm{sum}}$, $\chi_0^{(3)} = 2\,c\,\varepsilon_0\varepsilon_r n_{2,sum}$, and $n_{2,\mathrm{sum}} = (2/3)n_{2,el} + n_{2,c} + n_{2,l} + n_{2,d}$. The resulting values were $\chi_0^{(3)} = 3.81677\times10^{-20}\mathrm{m}^2/\mathrm{V}^2$, $f_{\mathrm{el}} = 0.0362$, $f_c = 0.0362$, $f_l = 0.2754$ and $f_d = 0.65217$.